\documentclass[fleqn,usenatbib]{mnras}

\usepackage[T1]{fontenc}

\DeclareRobustCommand{\VAN}[3]{#2}
\let\VANthebibliography\thebibliography
\def\thebibliography{\DeclareRobustCommand{\VAN}[3]{##3}\VANthebibliography}

\usepackage{graphicx}	
\usepackage{amsmath}	
\usepackage{amssymb}	

\usepackage{longtable}
\usepackage{lscape}
\usepackage{siunitx} 
\usepackage{threeparttable}
\usepackage{multirow}	
\usepackage{graphicx} 	
\usepackage{subcaption}
\usepackage{eso-pic}
\usepackage{mwe}

\newcommand{\mstellar}{\ensuremath{M_\mathrm{stellar}}}

\defcitealias{Wiseman2020}{W20}
\defcitealias{Smith2020}{S20}
\defcitealias{Kelsey2021}{K21}
\defcitealias{BroutScolnic2021}{BS21}

\newcommand{\fref}[1]{Figure~\ref{#1}}
\newcommand{\tref}[1]{Table~\ref{#1}}

\newcommand{\cref}[1]{Chapter~\ref{#1}}
\newcommand{\sref}[1]{Section~\ref{#1}}
\newcommand{\aref}[1]{Appendix~\ref{#1}}
\usepackage{newtxtext,newtxmath}

\usepackage{eso-pic}

\AddToShipoutPictureBG*{%
  \AtPageUpperLeft{%
    \hspace{0.75\paperwidth}%
    \raisebox{-4.5\baselineskip}{%
      \makebox[0pt][l]{\textnormal{DES-2021-0677}}
}}}%

\AddToShipoutPictureBG*{%
  \AtPageUpperLeft{%
    \hspace{0.75\paperwidth}%
    \raisebox{-5.5\baselineskip}{%
      \makebox[0pt][l]{\textnormal{FERMILAB-PUB-22-558-PPD}}
}}}%

\title[Concerning Colour]{Concerning Colour: The Effect of Environment on Type Ia Supernova Colour in the Dark Energy Survey}

\author[L.~Kelsey et al.]{
\parbox{\textwidth}{
\Large
L.~Kelsey,$^{1,2}$
M.~Sullivan,$^{2}$
P.~Wiseman,$^{2}$
P.~Armstrong,$^{3}$
R.~Chen,$^{4}$
D.~Brout,$^{5,6}$
T.~M.~Davis,$^{7}$
M.~Dixon,$^{8}$
C.~Frohmaier,$^{1,2}$
L.~Galbany,$^{9,10}$
O.~Graur,$^{1}$
R.~Kessler,$^{11,12}$
C.~Lidman,$^{13,3}$
A.~M\"oller,$^{8}$
B.~Popovic,$^{4}$
B.~Rose,$^{4}$
D.~Scolnic,$^{4}$
M.~Smith,$^{14}$
M.~Vincenzi,$^{4}$
T.~M.~C.~Abbott,$^{15}$
M.~Aguena,$^{16}$
S.~Allam,$^{17}$
O.~Alves,$^{18,16}$
J.~Annis,$^{17}$
D.~Bacon,$^{1}$
E.~Bertin,$^{19,20}$
S.~Bocquet,$^{21}$
D.~Brooks,$^{22}$
D.~L.~Burke,$^{23,24}$
A.~Carnero~Rosell,$^{25,16,26}$
M.~Carrasco~Kind,$^{27,28}$
J.~Carretero,$^{29}$
M.~Costanzi,$^{30,31,32}$
L.~N.~da Costa,$^{16}$
M.~E.~S.~Pereira,$^{33}$
S.~Desai,$^{34}$
H.~T.~Diehl,$^{17}$
S.~Everett,$^{35}$
I.~Ferrero,$^{36}$
J.~Frieman,$^{17,12}$
J.~Garc\'ia-Bellido,$^{37}$
D.~Gruen,$^{21}$
R.~A.~Gruendl,$^{27,28}$
J.~Gschwend,$^{16,38}$
G.~Gutierrez,$^{17}$
S.~R.~Hinton,$^{7}$
D.~L.~Hollowood,$^{39}$
K.~Honscheid,$^{40,41}$
D.~J.~James,$^{6}$
K.~Kuehn,$^{42,43}$
N.~Kuropatkin,$^{17}$
G.~F.~Lewis,$^{44}$
J. Mena-Fern{\'a}ndez,$^{45}$
R.~Miquel,$^{46,29}$
A.~Palmese,$^{47}$
F.~Paz-Chinch\'{o}n,$^{27,48}$
A.~Pieres,$^{16,38}$
A.~A.~Plazas~Malag\'on,$^{49}$
M.~Raveri,$^{50}$
M.~Rodriguez-Monroy,$^{45}$
A.~K.~Romer,$^{51}$
E.~Sanchez,$^{45}$
V.~Scarpine,$^{17}$
M.~Schubnell,$^{18}$
I.~Sevilla-Noarbe,$^{45}$
E.~Suchyta,$^{52}$
M.~E.~C.~Swanson,$^{53}$
G.~Tarle,$^{18}$
D.~L.~Tucker,$^{17}$
and N.~Weaverdyck$^{18,54}$
\begin{center} (DES Collaboration) \end{center}
}
\vspace{0.4cm}
\\
\parbox{\textwidth}{Affiliations are listed at the end of the paper}
}

\date{Accepted XXX. Received YYY; in original form ZZZ}

\pubyear{2022}

\begin{document}
\label{firstpage}
\pagerange{\pageref{firstpage}--\pageref{lastpage}}
\maketitle

\begin{abstract}

Recent analyses have found intriguing correlations between the colour ($c$) of type Ia supernovae (SNe Ia) and the size of their \lq mass-step\rq, the relationship between SN Ia host galaxy stellar mass ($\mstellar$) and SN Ia Hubble residual, and suggest that the cause of this relationship is dust. Using 675 photometrically-classified SNe Ia from the Dark Energy Survey 5-year sample, we study the differences in Hubble residual for a variety of global host galaxy and local environmental properties for SN Ia subsamples split by their colour. We find a $3\sigma$ difference in the mass-step when comparing blue ($c<0$) and red ($c>0$) SNe. We observe the lowest r.m.s. scatter ($\sim0.14$\,mag) in the Hubble residual for blue SNe in low mass/blue environments, suggesting that this is the most homogeneous sample for cosmological analyses. By fitting for $c$-dependent relationships between Hubble residuals and $\mstellar$, approximating existing dust models, we remove the mass-step from the data and find tentative $\sim 2\sigma$ residual steps in rest-frame galaxy $U-R$ colour. This indicates that dust modelling based on $\mstellar$ may not fully explain the remaining dispersion in SN Ia luminosity. Instead, accounting for a $c$-dependent relationship between Hubble residuals and global $U-R$, results in $\leq1\sigma$ residual steps in $\mstellar$ and local $U-R$, suggesting that $U-R$ provides different information about the environment of SNe Ia compared to $\mstellar$, and motivating the inclusion of galaxy $U-R$ colour in SN Ia distance bias correction.

\end{abstract}



\section{Introduction} \label{intro}

The improved standardisation of type Ia supernovae (SNe Ia) is important to constrain their luminosity dispersion and gain further understanding of the dark energy equation-of-state parameter, $w$. By applying corrections based on empirical relationships between their brightness and light-curve width \citep[the \lq{brighter-slower}\rq\ relation;][]{Rust1974, Pskovskii1977, Phillips1993} and their brightness and optical colour \citep[the \lq{brighter-bluer}\rq\ relation;][]{Riess1996, Tripp1998}, their luminosity dispersion can be reduced to $\sim0.14$\,mag \citep{Scolnic2018}. After accounting for observational uncertainties, $\sim0.08$--$0.10$\,mag of \lq intrinsic dispersion\rq\ remains \citep[e.g.][]{Brout2019}.

In addition to these traditional light-curve corrections, there are additional correlations between the corrected SN Ia luminosity and various host galaxy \lq environmental\rq\ properties. The most well-studied of these is the \lq{mass step}\rq\  \citep[e.g.][]{Sullivan2010,Kelly2010,Lampeitl2010,Gupta2011,Johansson2013,Childress2013,Uddin2017,Uddin2020,Smith2020,Ponder2020, Popovic2021}, in which SNe Ia in more massive galaxies are more luminous after corrections than their counterparts occurring in galaxies with lower stellar masses. This step is typically measured through differing average Hubble residuals\footnote{Hubble residual: the difference between the measured distance modulus ($\mu_\mathrm{obs}$) to each SN and the distance modulus calculated from the best-fit cosmology ($\mu_{\mathrm{cosmo}}$).} on either side of some division in environmental property, e.g. high and low stellar mass. The astrophysical reasons for this disparity are unclear, however it is known that the stellar mass ($\mstellar$) of a galaxy correlates with the stellar ages, gas-phase and stellar metallicities, and dust content \citep{Tremonti2004, Gallazzi2005,Garn2010,Bravo2011,Zahid2013}, suggesting that the trends between corrected SN Ia brightness and host stellar mass could be due to differences in intrinsic SN progenitor properties \citep[e.g., age or metallicity;][]{Timmes2003,Ropke2004,Kasen2009,Bravo2010} or dust \citep[e.g.,][]{BroutScolnic2021}, or both. The physical nature of the dominant underlying effect remains an open question.

In addition to looking at the $\mstellar$ of the galaxy, some studies \citep[e.g. ][]{Lampeitl2010,Sullivan2010,DAndrea2011,Childress2013,Pan2014, Wolf2016,Uddin2017,Kim2019,Kelsey2021} also consider other environmental properties such as the star formation rate (SFR), specific star formation rate (sSFR; SFR per unit $\mstellar$) or rest-frame colour (e.g. $U-R$). These properties are correlated with $\mstellar$; the most massive galaxies tend to be redder, more passive, with the lowest sSFR, whilst the lower mass galaxies tend to have more recent or ongoing star formation. These parameters provide other complementary ways to probe the stellar populations of the SN host galaxies, and may also provide insight into potential ages of the host stellar populations. Similarly sized SN luminosity steps have been found for global host galaxy sSFR, with $>3\sigma$ evidence that SNe Ia in low sSFR galaxies are brighter on average than those in higher sSFR galaxies after corrections. The most accurate tracer to determine the relationship between magnitude and environmental property for use in cosmology remains unclear \citep{Briday2021}.

Alongside the host galaxy correlations, a wealth of studies \citep{Rigault2013,Rigault2015,Rigault2020,Jones2015,Jones2018,MorenoRaya20162,MorenoRaya2016,Roman2018,Galbany2018,Rose2019,Kim2018,Kim2019,Kelsey2021} have shown that looking at the local region around the SN, rather than the global properties of the entire host galaxy, can provide a better understanding of the SN progenitor environment. Global galaxy properties are weighted by surface brightness, meaning that global measurements are most representative of the properties of the brightest galactic regions, and thus may not accurately describe the true environment of the progenitor and resulting SN \citep{Rigault2013}. For example, a SN Ia may be located within a locally passive region within a globally star-forming galaxy, or vice versa \citep[although local star-forming regions in globally passive galaxies are uncommon, e.g.,][]{Rigault2013}. \citet{Rose2021} suggest that combining corrections based on host galaxy stellar mass and local stellar age provides the best improvement to SN Ia standardisation at $>3\sigma$, reducing the unexplained scatter by $\sim 10$ per cent.. 

Recent analyses \citep{Brout2019,Smith2020,BroutScolnic2021,Kelsey2021} have shown that the magnitude of this step in average luminosity or Hubble residual with environmental property changes when considering SNe of different colours. \citet[hereafter \citetalias{Kelsey2021}]{Kelsey2021} found a significant ($\sim 3\sigma$) difference between the step sizes for subsamples comprised of \lq red\rq\ and \lq blue\rq\ SNe Ia, with bluer SNe (defined as having a SALT2 \citep{Guy2007,Guy2010} colour $c$ of $c < 0$) being more homogeneous and displaying no significant step, whilst the redder ($c>0$) SNe have a higher dispersion and larger step sizes. 

Analyses of the underlying relationships between SN Ia colour $c$ and the properties of their host galaxy environments have grown over the past year, with suggestion that the differing average Hubble residuals in low and high mass galaxies are caused by differences in dust properties for SNe with different $c$ \citep{BroutScolnic2021,Popovic2021,Popovic2022}. Bluer SNe ($c < 0$) will suffer less dust extinction \citep{Jha2007} and therefore less scatter from event-to-event than red ($c > 0$) SNe. The presence of dust along the line of sight reddens the SN by differing amounts dependent on the properties of the dust, and therefore may not be the same for all SNe Ia \citep{Gonzalez-Gaitan2020, Thorp2021}. There is known variation in the total-to-selective extinction ratio ($R_V$) along different lines of sight in the Milky Way \citep[e.g. ][]{Schlafly2016}, so logically $R_V$ should vary between, and even within, different SNe host galaxies. This is considered in \citet{Chen2022} for a sample of DES SNe Ia in redMaGiC galaxies, and \citet{Rose2022} for a sample of Pantheon+ SNe Ia \citep{Scolnic2021,Brout2022}. \citet{Meldorf2022} suggest that a correlation between host-$R_V$ and SNe Ia properties indicate that intrinsic scatter is driven by $R_V$.

An alternate explanation is that red and blue SNe Ia represent differing progenitor paths \citep[e.g. ][]{Milne2013, Stritzinger2018, Gonzalez-Gaitan2020, Kelsey2021}. Blue objects are considered to be comprised of one distinct set of progenitors (hence displaying no significant step in Hubble residual across hosts of differing masses), whilst red objects are likely a combination of different progenitors or explosion mechanisms (including the blue SNe that have been reddened by dust), causing a step in Hubble residual between different mass hosts to be observed. Environmental studies may find evidence for this by analysis of the stellar population age of the region surrounding the SNe. 

Regardless of the cause of the Hubble residual step, such studies indicate that blue SNe Ia, particularly those in bluer/low-mass environments, are more homogeneous and thus are better for use in cosmology \citep{Graur2015,Gonzalez-Gaitan2020, Kelsey2021}. In this study, we further examine the idea that blue SNe Ia are more homogeneous by studying the differences in Hubble residual for subsets of SNe Ia divided by SN colour, using photometrically-confirmed SNe Ia from the Dark Energy Survey (DES) SN programme (DES-SN) five-year cosmological sample. 

Our paper is structured in the following way. In \sref{data-and-methods}, we describe the DES-SN SN Ia sample that was used in this analysis and present the method to obtain environmental properties from photometric data. We discuss the results of our study in \sref{results} and \sref{colour}, and additional analysis in \sref{discussion}. Finally, in \sref{summary} we summarise and conclude. 

\section{Data and methods} \label{data-and-methods}

We begin by describing the SN Ia sample used in our analysis, and the methods used to obtain information about their galactic environments. 

\subsection{The DES-SN photometric SN Ia sample}

DES is an optical imaging survey that uses four independent astrophysical probes to measure the properties of dark energy \citep{DES2016}. Here we use a sample of SNe Ia discovered by the dedicated SN programme in DES, DES-SN \citep{Abbott2019}, comprised of SNe Ia discovered in imaging data acquired by the Dark Energy Camera \citep[DECam;][]{Flaugher2015}, mounted on the Blanco 4-m telescope at the Cerro Tololo Inter-American Observatory. The DES-SN programme was optimised for the detection of SNe Ia over the redshift range $0.2 < z < 1.2$ \citep{Bernstein2012,SmithDAndrea2020} for use in cosmology, observing ten 3-deg$^2$ fields with an average cadence of 7 days in four filters ($griz$). Our sample is taken from the full five years of the survey.

This sample differs from the DES-SN three-year (DES-SN3YR) sample used in \citetalias{Kelsey2021}: it includes data from the full five years of the survey instead of only the first three years \citep{Brout2019a}, and it includes both spectroscopically-confirmed and photometrically-classified SNe Ia where the redshift for each SN is determined by a spectroscopic redshift measurement of its host galaxy. The photometry is obtained using \texttt{diffimg} \citep{Kessler2015}. DES photometric classification is outlined in \citet{Vincenzi2020} and \citet{Moller2022}, with host association details in \citet{Wiseman2020}. As detailed in \citet{Vincenzi2020}, sample contamination from  core-collapse SNe and peculiar SNe Ia ranges from 0.8–3.5 per cent in DES, meaning that it has a negligible effect and is not a limiting uncertainty in DES cosmological analyses, such as this.

\subsubsection{SN Ia light-curve parameters} \label{params}

We use the SALT2 SN Ia light-curve model \citep{Guy2007,Guy2010} to fit the SN Ia light curves and obtain estimates of their \lq stretch\rq\ ($x_1$), \lq colour\rq\ ($c$) and $m_B$ ($-2.5\log(x_0)$, where $x_0$ is the fitted amplitude). SALT2 is trained with the JLA compilation SN sample, and implemented in the \textsc{snana} software package \citep{Kessler2009}. In this analysis, we use 1D BBC bias corrections \citep{KesslerScolnic2017} to correct for selection bias with a \citet{Guy2010} intrinsic scatter model (consistent with \citetalias{Smith2020} and \citetalias{Kelsey2021}). We do not employ BEAMS, instead setting each P(Ia) = 1. The light curve parameters are used to calculate Hubble residuals:
\begin{equation}
    \Delta\mu = \mu_\mathrm{obs} - \mu_\mathrm{cosmo},
\label{equation:resi}
\end{equation}
where $\mu_\mathrm{cosmo}$ is the fixed distance modulus calculated from a reference cosmology (flat $\Lambda CDM$ with $w = -1$), and $\mu_\mathrm{obs}$ is the measured distance modulus \citep[e.g.,][]{Tripp1998,Astier2006}:
\begin{equation}
    \mu_\mathrm{obs} = m_B - M_0 + \alpha x_1 - \beta c + \mu_\mathrm{bias},
\label{equation:hubble}
\end{equation}
with $\alpha$, $\beta$ and $M_0$ as nuisance parameters describing the SN population in the BBC fit.

The $\mu_\mathrm{bias}$ represents a correction that is applied to each SN to account for survey selection effects. This correction is typically either a \lq 1D correction\rq\ as a function of redshift, or a \lq 5D correction\rq\ as a function of \{$z, x_1, c, \alpha, \beta$\} \citep{KesslerScolnic2017}. The 1D correction does not account for the $c$-dependent selection bias (bluer SNe are brighter and easier to observe), which results in a trend of $\Delta\mu\ \textrm{vs}\ c$ for blue SNe. A discussion of the differences between 1D and 5D corrections with regards to host galaxy correlations in the DES-SN3YR sample can be found in \citet[hereafter S20]{Smith2020}. We consider a 5D bias correction in \aref{BBC5D}, finding no significant difference in our results. 

In cosmological analyses, there is an additional host galaxy \mstellar\ correction added to Eq. \ref{equation:hubble}, $\gamma G_\textrm{host}$, where the nuisance parameter $\gamma$ is analogous to $\alpha$ and $\beta$, and $G_\textrm{host}$ is a step function typically located at $\log (\mstellar / M_\odot) = 10$. We do not use this additional correction in our analysis, as we want to study the overall cause of the additional dispersion and determine if it can be explained with this simple correction.  

We assume a spatially-flat $\Lambda$CDM model, with a matter density $\Omega_M = 0.3$ and Hubble constant $H_0 = 70$\,km\,s$^{-1}$\,Mpc$^{-1}$ as a reference cosmology for the calculation of $\Delta\mu$.

\subsection{SN host galaxy photometry}

Here we briefly describe the DES-SN image stacking procedure and methods used to obtain photometry of each SN Ia host galaxy and region local to the SN event. The method is identical to that used in \citetalias{Kelsey2021}, and details can be found therein. 

Host galaxies are assigned using the directional light radius method \citep[DLR;][]{Sullivan2006,Gupta2016} and are catalogued in \citet{Wiseman2020}. The DLR is a measure of the separation distance between the SN and each galaxy, normalised by the apparent size of the galaxy light profile being considered  \citep[obtained from high-quality depth-optimised coadded images;][]{Wiseman2020}, in terms of the elliptical radius along a line connecting the SN to the host center.

The majority of host galaxy spectroscopic redshifts for the DES photometric sample were provided by the OzDES programme \citep{Yuan2015, Childress2017, Lidman2020} using the Anglo-Australian Telescope (AAT). A subset of host galaxy redshifts were obtained from external catalogues of prior surveys that overlapped with the DES-SN fields. Details of host galaxy association and redshifts can be found in \citet{Vincenzi2020, Moller2022}.   

We use the \lq seeing-optimised\rq\ DES image stacks described in \citetalias{Kelsey2021} \citep[created following][]{Wiseman2020}. Single-epoch exposures are added to the stack if they pass given quality cuts; for this analysis we use exposures with a $\tau$ (ratio between effective exposure time and true exposure time) of $>$ 0.02 and a point spread function (PSF) full-width half-maximum (FWHM) $< 1.3\arcsec$\ in all filters. This provides a balance between seeing and redshift coverage for our analysis.  

Following \citetalias{Kelsey2021}, photometry for the host galaxy (\lq global\rq\ photometry) is measured using  \textsc{Source Extractor} \citep{Bertin1996} on the stacked $griz$ images. 

We also measure local photometry at the SN position using a 4\,kpc aperture radius following \citetalias{Kelsey2021}, based on the quality of our stacked images. Local aperture photometry is measured using \textsc{aperture\_photometry} from the \textsc{photutils} Python module \citep{Bradley2019}, and photometric uncertainties are calculated (also using this tool) using the root-mean-square of the background-subtracted stacked images as an error map.

All our measured $griz$ fluxes are corrected for Milky Way dust extinction using colour excess $E(B-V)$ values from \citet{Schlegel1998} and multiplicative coefficients for the DES filters of $R_{g} = 3.186$, $R_{r} = 2.140$, $R_{i} = 1.569$ and $R_{z} = 1.196$ \citep{Abbott2018}, calculated using a Fitzpatrick reddening law \citep{Fitzpatrick1999}.

\subsection{SN host galaxy SED fitting}

As per \citet{Smith2020,Wiseman2020,Kelsey2021}, we use spectral energy distribution (SED) fitting and templates based on the \textsc{p\'egase} spectral evolution code \citep{Fioc1997,Fioc2019} assuming a \citet{Kroupa2001} initial mass function (IMF) and a series of 9 smooth exponentially-declining star-formation histories, each with 102 time steps, in order to estimate the physical parameters from the photometry. Synthetic DES photometry is generated for each SED  template and, using $\chi^2$ minimisation, is matched with the measured photometry for each region \citep[e.g.,][]{Wiseman2020}. We apply a foreground dust screen with $E(B-V) = 0$ to $0.3$\,mag in steps of $0.05$ to account for dust extinction, and only consider solutions younger than the age of the universe for each SN redshift. 

From this SED fitting we obtain the stellar mass ($\mstellar$, in $\mathrm{M}_{\sun}$) and the rest-frame $UBVR$ magnitudes for all global and local regions. For each set of photometry we additionally use a Monte Carlo process adjusting the observed photometry according to its uncertainties, with 1000 iterations in order to estimate the uncertainties in the above parameters. Full details of this process can be found in \citetalias{Smith2020}. 

As described in \citetalias{Kelsey2021}, we apply a \lq mangling\rq\ \citep{Hsiao2007,Conley2008} correction, adjusting the best-fitting SED for each host galaxy or region using a wavelength-dependent spline multiplicative function to ensure that the SED exactly reproduces the observed photometry. This procedure allows rest-frame $UBVR$ magnitudes to be accurately calculated.  

As in \citetalias{Kelsey2021}, we focus our analysis on rest-frame $U-R$. We choose this colour because it spans the greatest wavelength range in our observer-frame ($griz$) photometry (above our redshift cut, discussed in \sref{selection}, we lose rest-frame $R$ band), it is an approximate tracer of the SFR, it carries some age information of the galaxy \citep{Trayford2016}, and it correlates with galaxy morphology \citep[correlation with $u-r$;][]{Strateva2001,Lintott2008}. By assuming that the difference in SN luminosities is due to local stellar population age, rest-frame $U-R$ has been shown to be the best photometric tracer of this parameter \citep{Briday2021}, making it highly suitable for high redshift cosmology, where spectroscopy may not be obtained of each SN environment. Furthermore, recent analyses suggest that combining environmental corrections such as global host galaxy stellar mass, and local age (potentially with colour as a proxy) may provide the best standardisation for SNe cosmology \citep{Rose2019,Rose2021,Rigault2020}.

\subsection{SN selection requirements} \label{selection}

From the SuperNNova classifier, we require each candidate SN Ia to have a probability of being a SN Ia of P(Ia) $> 0.5$.\footnote{See \aref{diff_class} for a discussion of the use of different classifiers, templates and $P(Ia)$ selection cuts.} We apply a redshift cut of $z < 0.6$, which ensures that at all redshifts the aperture size is larger than the smallest useful aperture ($\sigma$) of 0.55\arcsec\ for a maximum full-width half-maximum of 1.3\arcsec\ when approximating to a Gaussian following \citetalias{Kelsey2021}. Additionally, this redshift cut minimises selection biases, particularly in the shallow fields \citep{Kessler2019}. We apply a typical \lq{JLA-like}\rq\ \citep{Betoule2014} light-curve selection in $x_1$ and $c$, and their associated uncertainties. Based on the sample defined by \citet{Moller2022}, 787 SNe Ia pass these classification, redshift, and JLA-like cosmological cuts, including our additional constraint of $\sigma_c<0.1$ to match \citet{BroutScolnic2021}, and pass BBC 1D bias corrections. We apply a selection cut on $\sigma_{(U-R)} < 1$\,mag for both the global and local measurements to have well-constrained rest-frame $U-R$ colours. This cut also removes objects with large uncertainties in \mstellar\ and SFR. We also require viable SED-fits for global and local environments, finding that some local environments have too low signal-to-noise to be able to adequately fit a model to the SED. We further require that the uncertainties in derived environmental parameters were small ($\sigma_{(U-R)} < 1$ and $\sigma_{(\mstellar)} < 1$), and that the global mass obtained for an environment was greater than the local mass. A summary of the selection requirements applied is presented in \tref{table:5yrselection}. Using a BBC fit (Section~\ref{params}), we obtain values of $\alpha = 0.158\pm0.007$ and $\beta = 2.88\pm0.07$ for these data.
\begin{table}
\begin{center}
\caption{SN Ia sample selection.}
\begin{threeparttable}
\begin{tabular}{l c} 
\hline
Cut & Number of SNe Ia\\
\hline
Cosmology \tnote{1}\ \ \& $z < 0.6$ \& P(Ia)$ > 0.5$ & 787 \\
Local photometry SNR cut\tnote{ 2} & 705 \\
$\sigma_{(U-R)} < 1$ and $\sigma_{(\mstellar)} < 1$ \tnote{ 3} & 695 \\
$(\mstellar)_\textrm{global} \ge (\mstellar)_\textrm{local}$ & 675 \\
\hline
\end{tabular}
\begin{tablenotes}
\item[1] Cosmology refers to the JLA-like requirements on colour and stretch, i.e.: $|c|<0.3$, $\sigma_c<0.1$, $|x_1| < 3$, and $\sigma_{x_1}<1$. We also require $|\Delta\mu|/\sigma_{\mu} < 4$ which removes $4\sigma$ outliers in the Hubble residual, and for the SNe to pass 1D BBC bias corrections.
\item[2] This encompasses those lost due to poor quality local photometry, in which there was not enough signal-to-noise for the SED to obtain a suitable fit.
\item[3] These restrictions hold for both global and local properties.
\end{tablenotes}
\end{threeparttable}
\label{table:5yrselection}
\end{center}
\end{table}

We present the global and local photometry and derived environmental properties for the 675 SNe Ia used in this analysis in the online supplementary material. Light curves and associated environmental data for the full DES5YR sample will be released online with the cosmology data release.

\section{Global and Local Environments} \label{results}

\subsection{Environmental properties and colour}

\begin{figure*}
    \begin{center}
    \includegraphics[width=.45\textwidth]{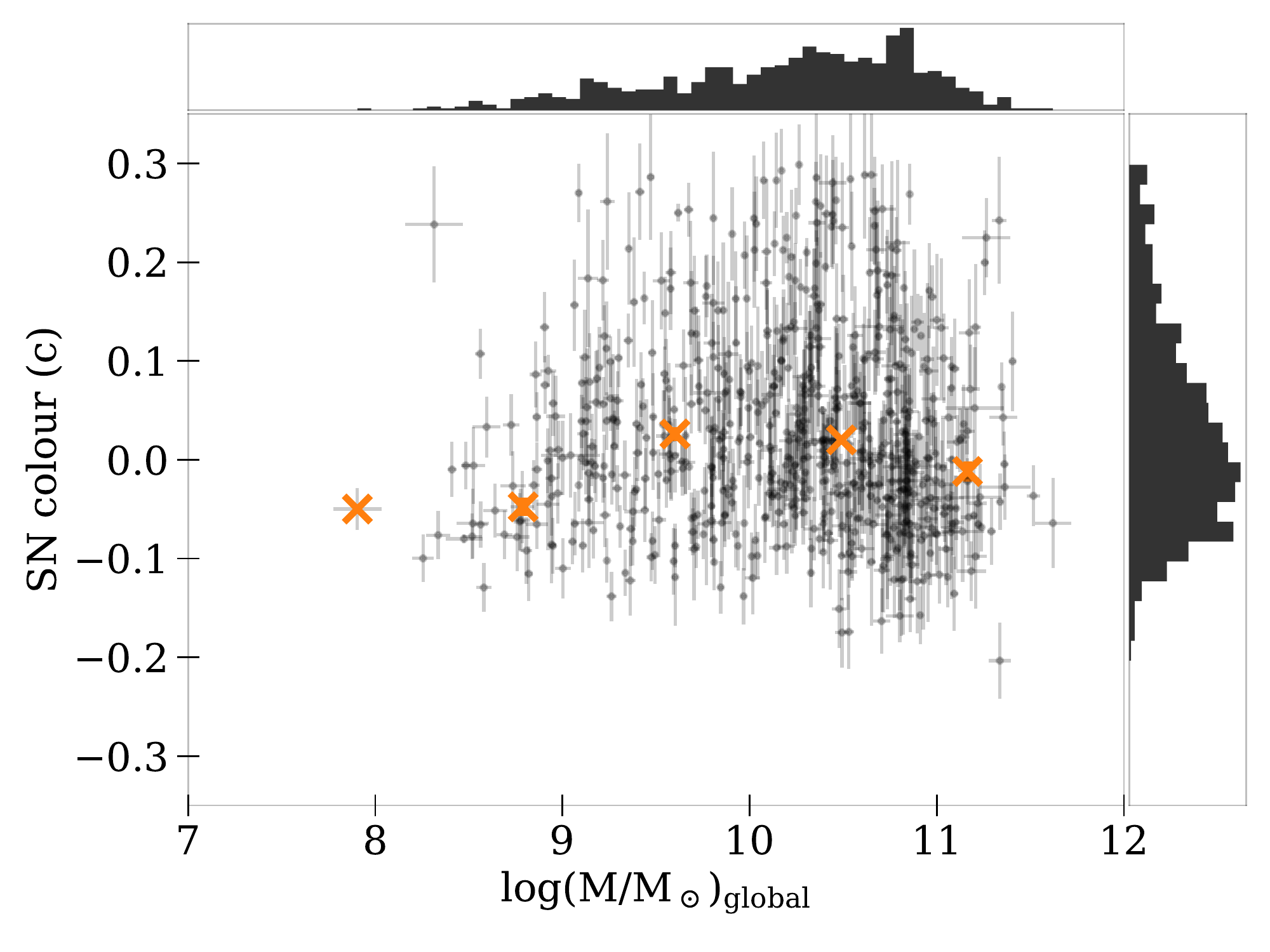}
    \includegraphics[width=.45\textwidth]{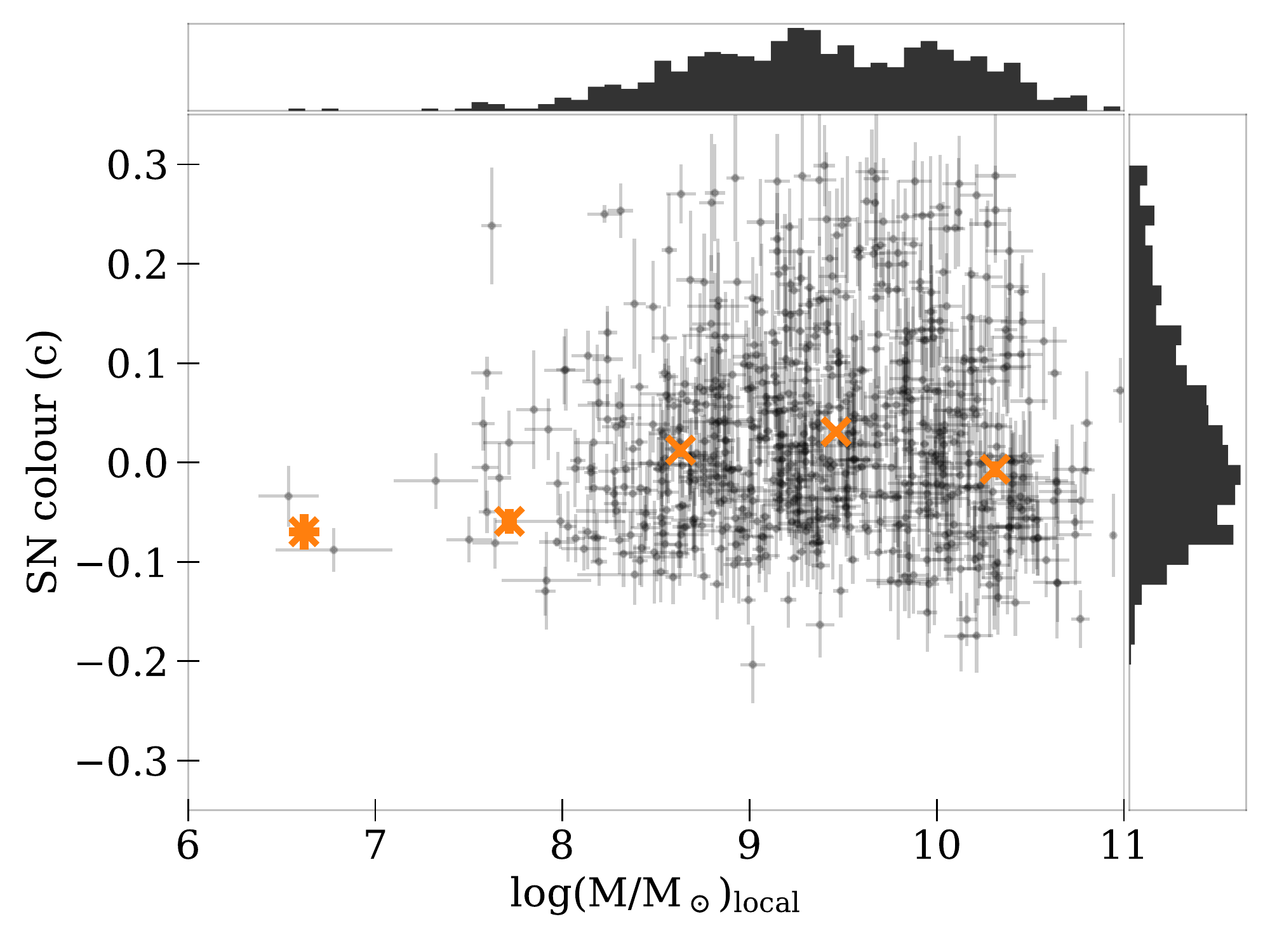}
    \includegraphics[width=.45\textwidth]{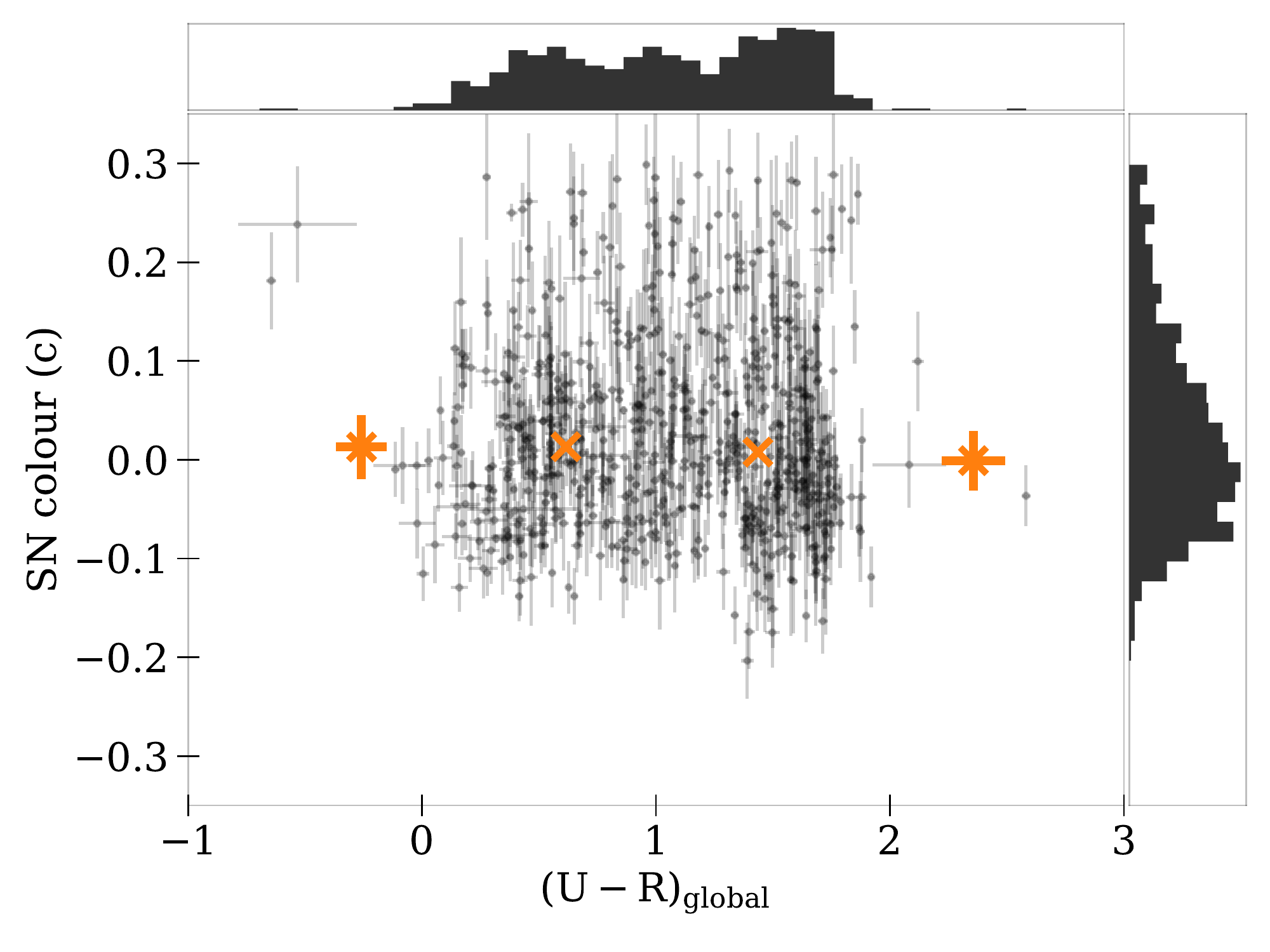}
    \includegraphics[width=.45\textwidth]{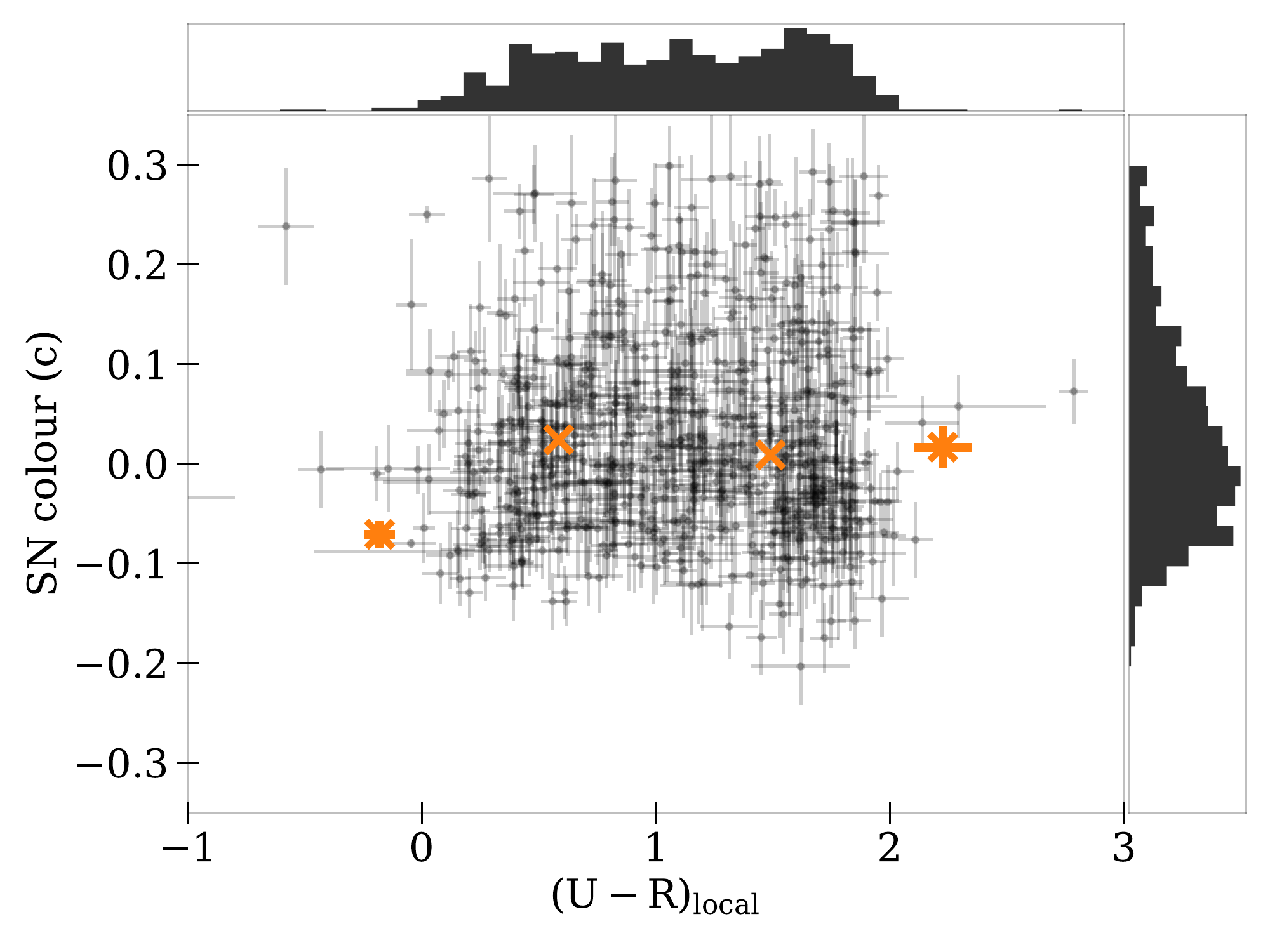}
    \caption{SN colour $c$ as a function of both global and local environmental properties: \mstellar\ (upper panels) and rest-frame $U-R$ (lower panels). The orange points show the binned weighted mean colours. The histograms show the distributions of $c$ and the environmental properties.}
    \label{fig:c_envprop_relations}
    \end{center}
\end{figure*}

We study the relationships between SN Ia colour and the SN Ia environmental properties \mstellar\ and rest-frame $U-R$ colour, both globally for the host galaxy, and for the \lq local\rq\ 4\,kpc radius regions. From Fig.~\ref{fig:c_envprop_relations}, trends in environmental properties with $c$ are shown, with more massive, redder galaxies and environmental regions hosting redder SNe Ia, consistent with prior (but weak) observed trends \citep{Sullivan2010,Childress2013,Kelsey2021, BroutScolnic2021, Popovic2021}. As in \citetalias{Kelsey2021}, we observe an absence of red SNe Ia in low-mass galaxies and, to a lesser extent, in bluer $U-R$ regions. This is not unexpected: the distribution of SN reddening is similar in low- and high-mass galaxies \citep{Popovic2022}, but the number of SNe in low mass galaxies (particularly below $\log(M/M_{\odot})<9.5$ where the effect is noticeable) is smaller, so the red tail of the distribution is simply not well sampled.

\subsection{Environmental Property steps in Hubble residual} \label{envHR}

\begin{figure*}
\begin{center}
\includegraphics[width=.49\linewidth]{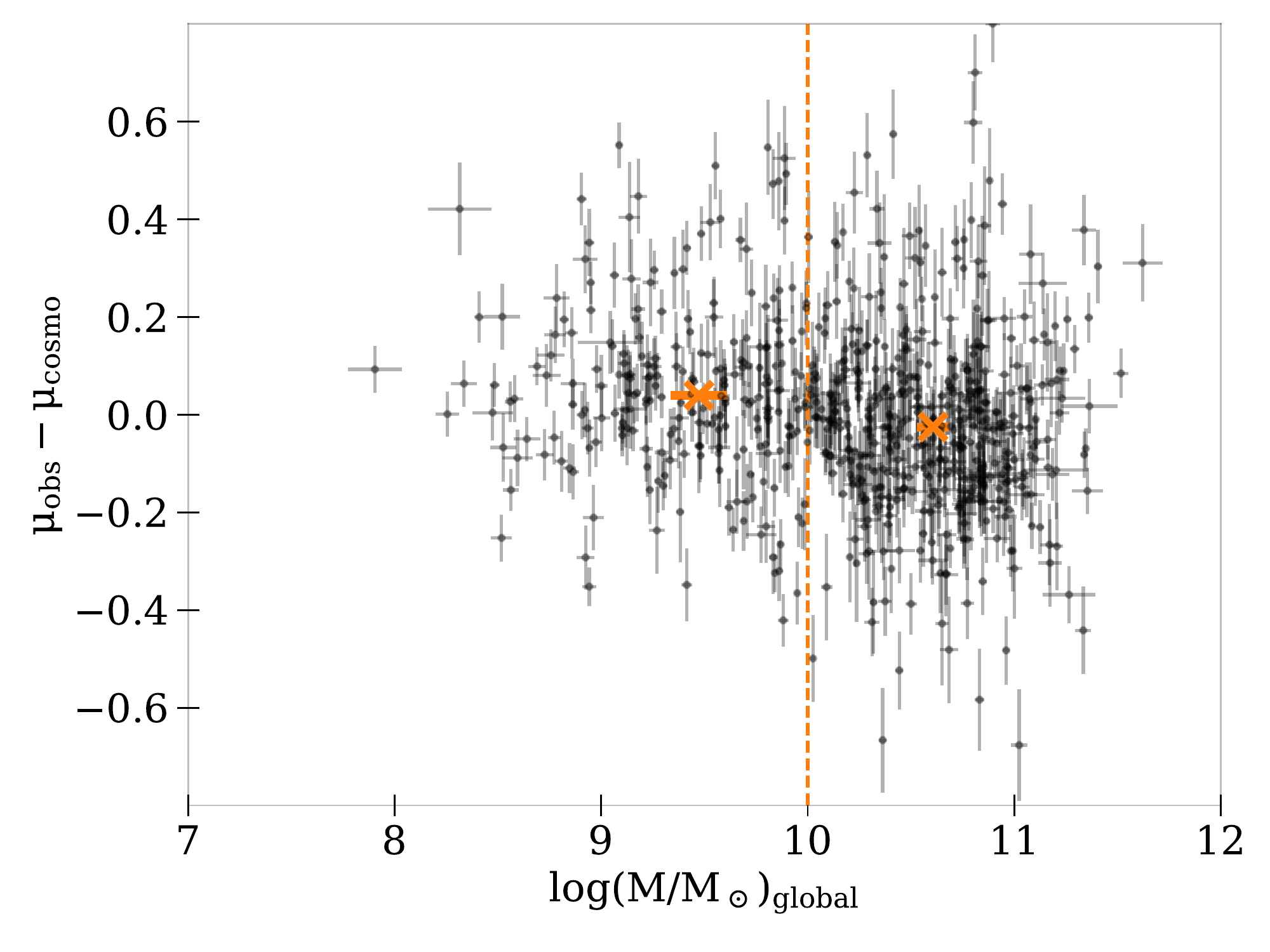}
\includegraphics[width=.49\linewidth]{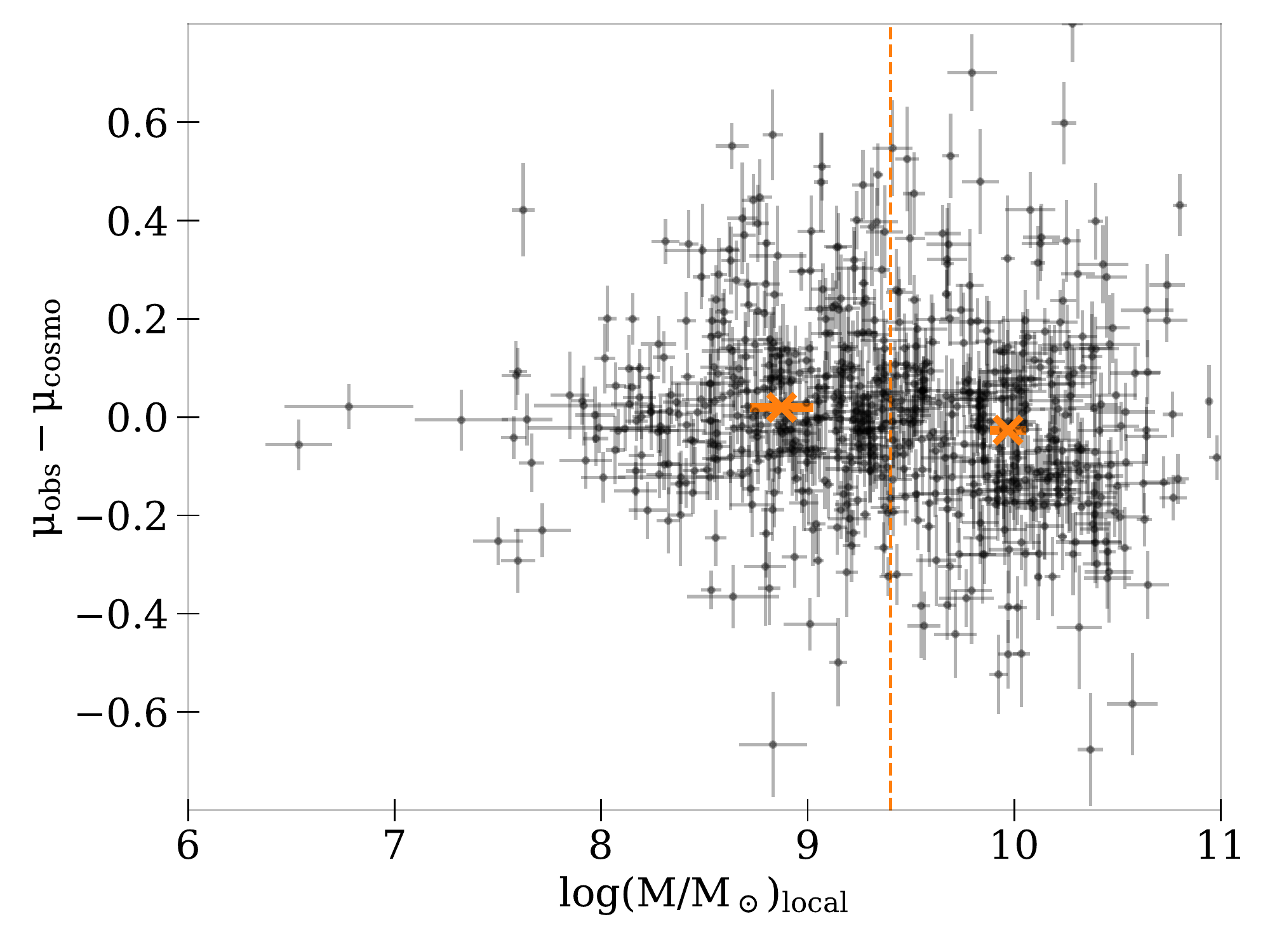}
\includegraphics[width=.49\linewidth]{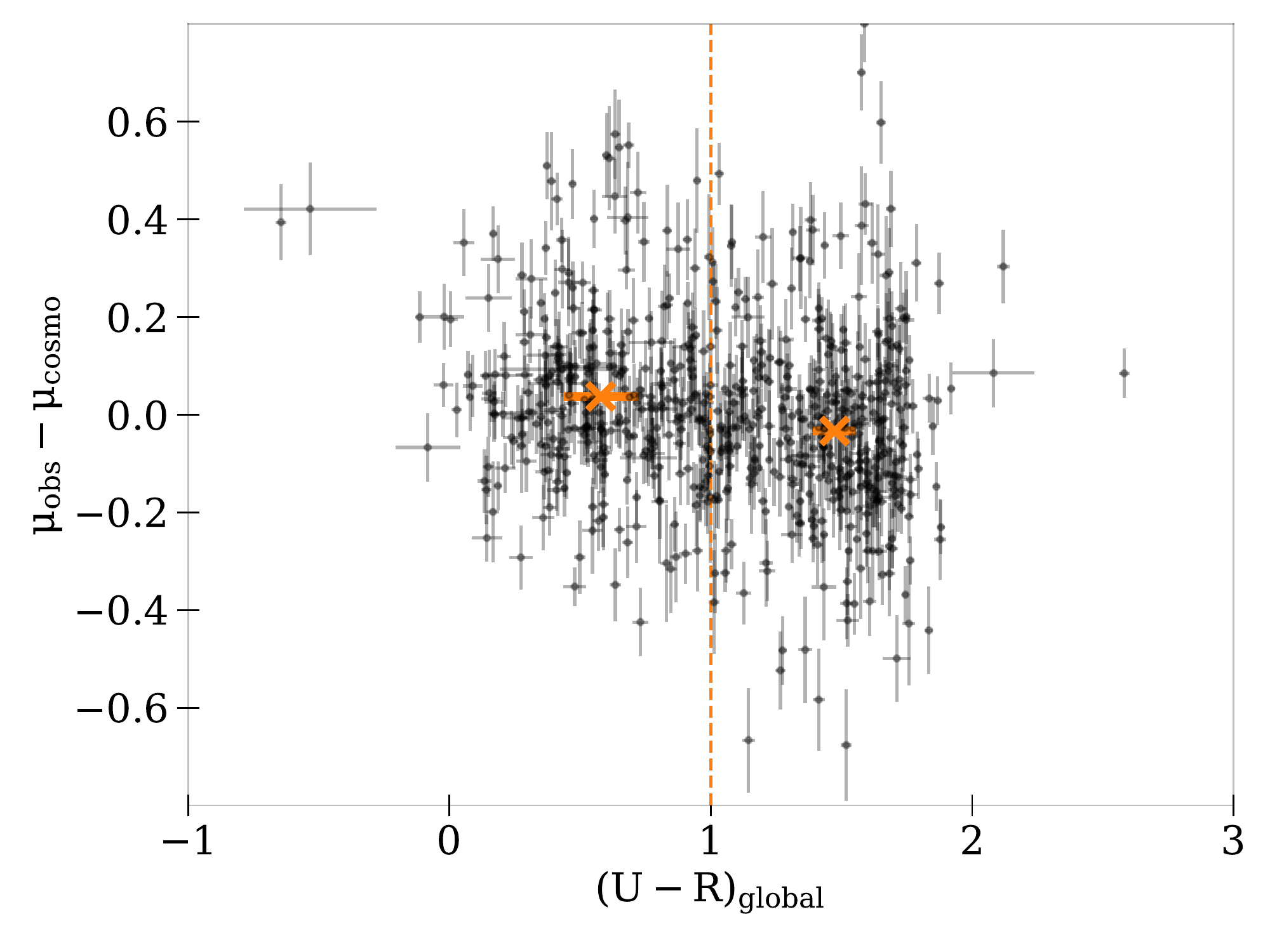}
\includegraphics[width=.49\linewidth]{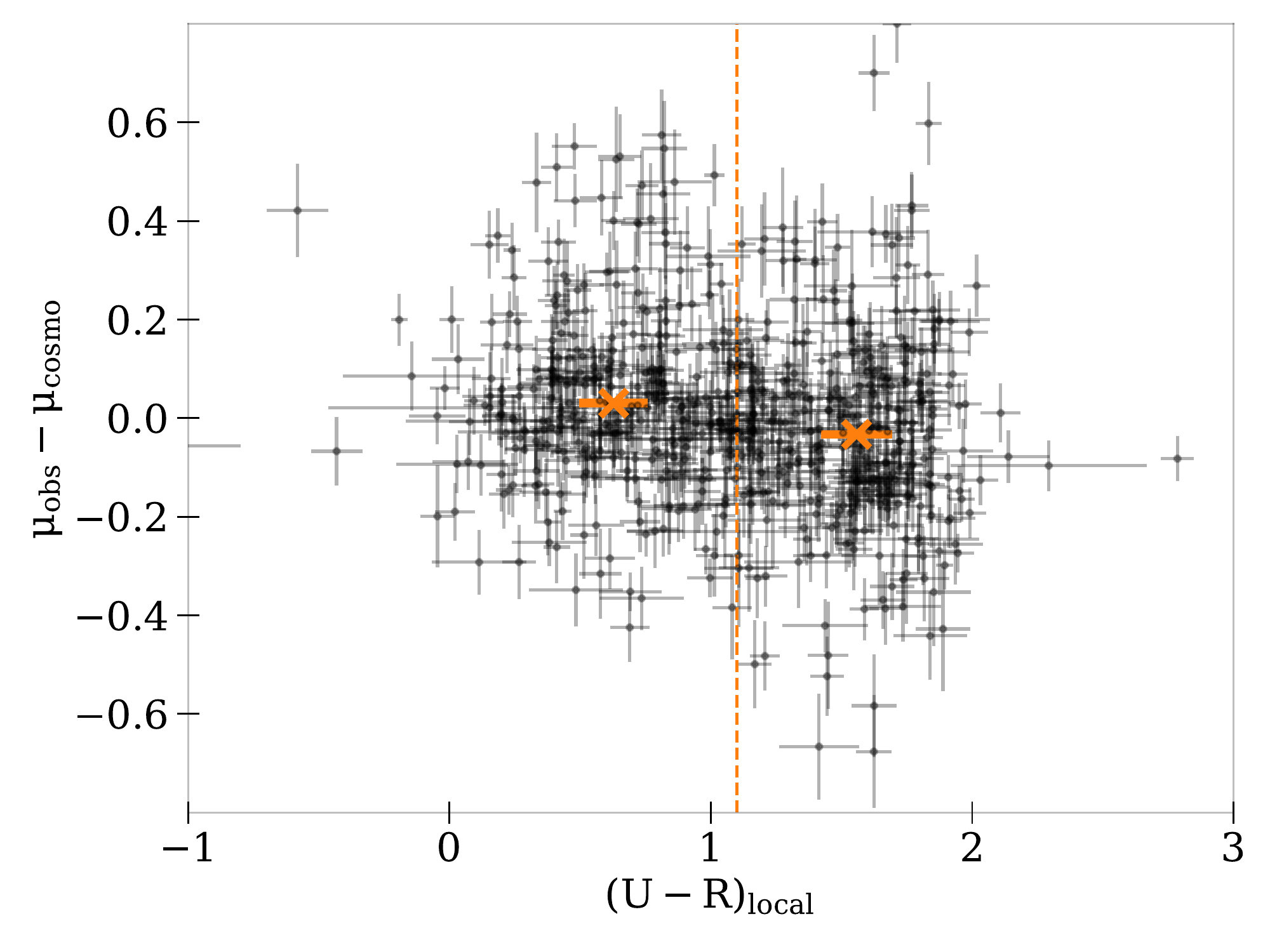}
\caption{Hubble residual plots for DES5YR as a function of (from top to bottom and left to right): global $\mstellar$, local $\mstellar$ within the 4\,kpc radius aperture, global rest-frame $U-R$ colour and local rest-frame $U-R$ colour. The orange dashed line is the split location, and the orange markers represent the bin means, displaying the steps in Hubble residual.}
\label{fig:5yr-hubble_mass_colour_step-litsplit}
\end{center}
\end{figure*}

We now turn to investigating the relationships between SN Ia Hubble residuals with $\mstellar$ and rest-frame $U-R$ colour, both globally and locally, for our DES5YR sample. We plot the Hubble residual vs. the chosen environmental property split into two bins at a chosen division point in \fref{fig:5yr-hubble_mass_colour_step-litsplit}, and measure the mean and dispersion in Hubble residual either side of this division. The magnitude of the \lq{step}\rq\ is simply the difference between the two means, provided with the statistical significance ($N\sigma$) of the difference. The resulting steps are presented in \tref{table:5yr_4values_BBC1D}.
We present the step values calculated with the following step locations (division points):

\begin{itemize}
    \item $\log (\mstellar/M_\odot)_{\textrm{global}} = 10.0$ \citep[e.g.,][]{Sullivan2010}
    \item $\log(\mstellar/M_\odot)_{\textrm{local}} = 9.4$ (this value represents the median local $\mstellar$ for this DES5YR sample)
    \item $(U-R)_{\textrm{global}} = 1.0$ \citepalias{Kelsey2021}
    \item $(U-R)_{\textrm{local}} = 1.1$ (this value represents the median local $U-R$ of this DES5YR sample, which is redder than that for \citetalias{Kelsey2021})
\end{itemize}  

\begin{table*}
\begin{center}
\caption{Hubble residual steps, and r.m.s. scatter in Hubble residual, across both stellar mass and $U-R$ for our DES5YR sample using a 1D bias correction.}
\begin{threeparttable}
\begin{tabular}{c c c c c c}

\hline
\multicolumn{2}{c}{Property} &  \multicolumn{2}{c}{Hubble Residual Step} & \multicolumn{2}{c}{Hubble Residual r.m.s.}\\ Name & Division Point & Magnitude & Sig. ($\sigma$)\tnote{b} & $<$ DP\tnote{c} & $>$ DP \\
\hline
Global Mass\tnote{a} & 10.0 & 0.065$\pm$0.013 & 4.91 & 0.186$\pm$0.017 & 0.194$\pm$0.013\\
Local Mass & 9.4 & 0.046$\pm$0.013 & 3.67 & 0.177$\pm$0.013 & 0.205$\pm$0.016\\
\hline
Global U-R & 1.0 & 0.071$\pm$0.012 & 5.72 & 0.181$\pm$0.015 & 0.199$\pm$0.015\\
Local U-R & 1.1 & 0.063$\pm$0.012 & 5.10 & 0.183$\pm$0.014 & 0.198$\pm$0.015\\
\hline
\end{tabular}
\begin{tablenotes}
\item[a] Mass in $\log (\mstellar / M_\odot)$
\item[b] Significance is quadrature sum.
\item[c] DP refers to the \lq{Division Point}\rq\ location of the environmental property step. For example, \lq{$<$DP}\rq\ indicates the lower mass or bluer environments. 
\end{tablenotes}
\end{threeparttable}
\label{table:5yr_4values_BBC1D}
\end{center}
\end{table*}

\citetalias{Kelsey2021} found that the majority (56\%) of SNe in the DES3YR spectroscopically-confirmed sample were located in local regions that were bluer than their host galaxy average, with a median local $U-R = 0.95$. Using this larger sample of DES5YR photometrically-classified SNe we find a different result, $62\%$ of SNe are located in regions that are locally redder than their host galaxy. This relationship is not redshift dependent, so is likely a feature of types of galaxies that are found to host SNe Ia in a photometric survey. For example, DES5YR has more SNe in high mass hosts than DES3YR, and contains many more SNe with particularly low DLR measurements. For DES3YR, which was spectroscopically-confirmed, SN spectra were required, which are more difficult to obtain for SNe near the centre of their host galaxies. For DES5YR, which was photometrically-identified, spectra from the SNe themselves were not needed, which removes a potential bias against SNe closer to the centre of their hosts.

\begin{figure}
\begin{center}
\includegraphics[width=\linewidth]{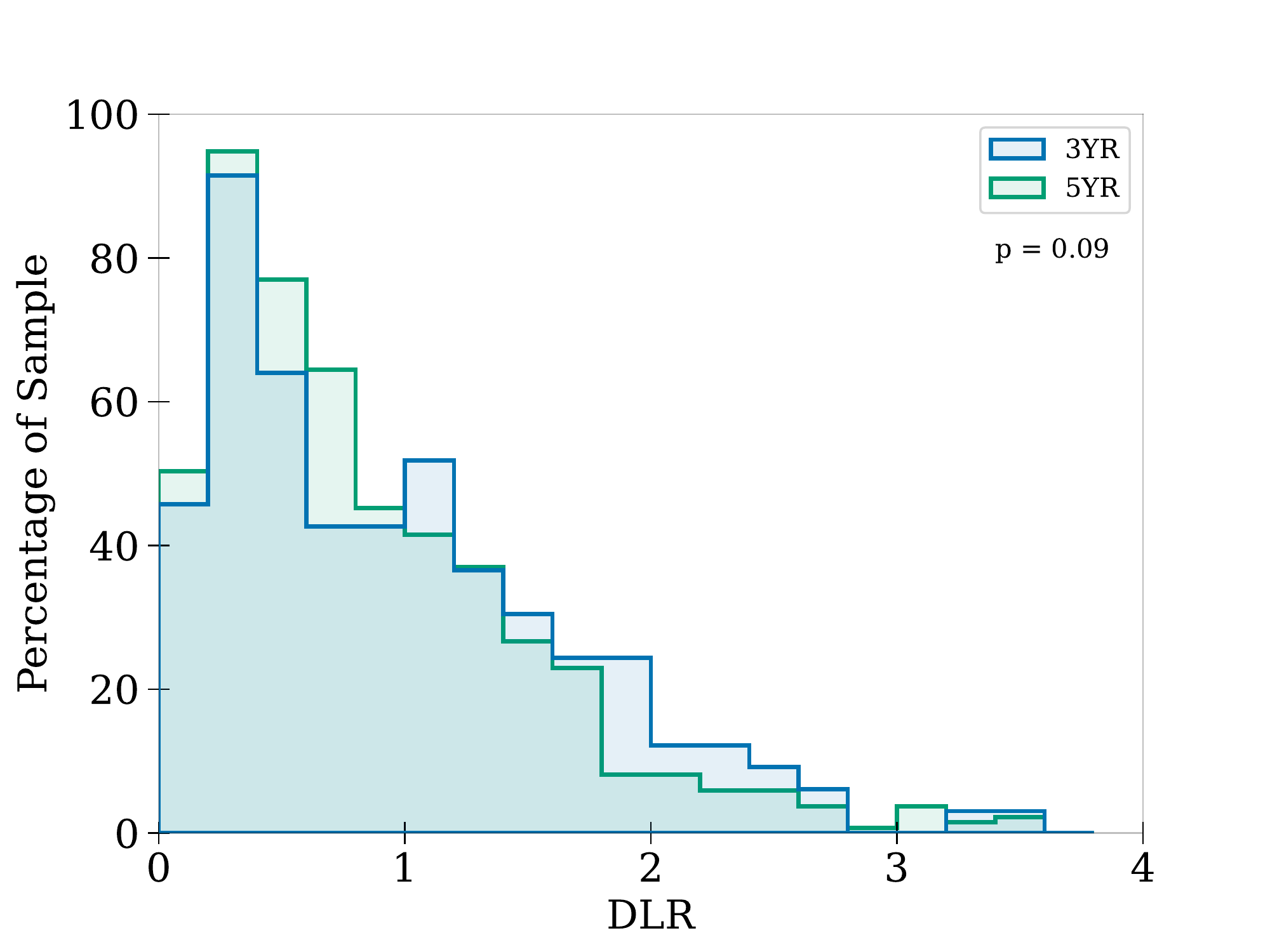}
\caption{Distribution of DLR values for DES3YR and DES5YR, given as a percentage of individual sample sizes (DES3YR = 164, DES5YR = 675 SNe Ia), showing that a greater percentage of SNe in DES5YR are closer to the center. The mean DLR for DES3YR = 0.96, and for DES5yr = 0.86. P-value ($p$) $= 0.09$ from KS testing is displayed in the top right corner.}
\label{fig:histDLR}
\end{center}
\end{figure}

By comparing the local $U-R$ to the DLR for DES5YR, the majority of SNe Ia that are in locally redder regions than their host galaxy average are located closer to the centre of their host galaxy. This is likely due to the colour gradients in galaxies in which elliptical and spiral galaxies are redder in the centre, getting progressively bluer outwards \citep[e.g.][]{Tortora2010}. This is an age effect, known as the \lq{inside-out scenario}\rq \citep[e.g.][]{Perez2013}. Star formation happens close to the galaxy centre, and over time is triggered towards the outskirts, generating an age gradient. This physical age gradient is observed in our data as a colour gradient. This colour gradient means that the average colour of a galaxy may be bluer than the colour of the central region.  Without SN spectra in DES5YR, more SNe Ia in the centres of galaxies are present in the sample as shown in \fref{fig:histDLR}, meaning that the effect of this colour gradient is more noticeable than for DES3YR. This in turn means that the median local $U-R$ is redder than for \citetalias{Kelsey2021}, and the median local $\mstellar$ is higher, motivating our choice of division point locations for local properties. 

\subsubsection{$\mstellar$}

Focusing first on global $\mstellar$, as presented in \tref{table:5yr_4values_BBC1D}, the Hubble residual step of $0.065\pm0.013$ mag ($4.9\sigma$) agrees with prior analyses \citep[e.g.][]{Sullivan2010, Childress2013, Smith2020}. For local $\mstellar$, the Hubble residual step is smaller ($0.046\pm0.013$) in magnitude, but is still $3.7\sigma$ in significance. For both global and local $\mstellar$, the r.m.s. values are lower for lower mass regions than for higher masses, consistent with \citetalias{Kelsey2021}. 

\subsubsection{$U-R$}

Moving to $U-R$, \tref{table:5yr_4values_BBC1D} shows that all the steps, for both local and global, are slightly smaller than, but consistent with the findings of \citet{Roman2018}, \citet{Rigault2020} and \citetalias{Kelsey2021}. The global colour step ($0.071\pm0.012 \textrm{ mag}; 5.7\sigma$) is larger than the stellar mass steps, but the local $U-R$ ($0.063\pm0.012 \textrm{ mag}; 5.1\sigma$) is consistent with the global mass. Overall, the $U-R$ colour steps are fairly similar whether measured globally or locally, in agreement with \citetalias{Kelsey2021} and \citet{Roman2018} (for $U-V$), likely due to the strong correlations between global and local colour. As in \citetalias{Kelsey2021} and as for the stellar masses, the r.m.s. values are lower in bluer environments.

We note that there appear to be potential outliers with particularly low or high Hubble residual values of $|\mu_\mathrm{obs} - \mu_\mathrm{cosmo}|>0.4$, which may indicate potential core-collapse contamination. To investigate this, we remove those objects and recalculate our environmental property steps in \tref{table:5yr_4values_BBC1D}, finding reductions in magnitudes of $<0.01$ mag (therefore less than the uncertainties). This also reduces the values of the r.m.s. on each side of the division points, but the overall trends remain consistent with lower scatter for the lower mass or bluer regions, agreeing with our results and those of \citetalias{Kelsey2021}. To avoid unwanted selection bias, we leave these objects in our sample.

\subsection{Refitting $\alpha$ and $\beta$}\label{refit}

In \citetalias{Kelsey2021}, a tentative $\sim2\sigma$ difference in optimal $\alpha$ and $\beta$ values on each side of the environmental property division point was found. To compare with DES5YR, we refit $\alpha$ and $\beta$ for subsamples split by $\mstellar$ and rest-frame $U-R$. This comparison could uncover whether the steps in luminosity are driven by underlying relationships between $x_1/c$ and environmental properties.

As can be seen in \tref{table:env_split-alpha-beta}, the differences in $\beta$ between subsamples with different environmental properties are the most pronounced, on the order of $3\sigma$ for all properties, with lower $\beta$ values for high mass or redder regions. This difference agrees with \citet{Sullivan2011, BroutScolnic2021, Kelsey2021}, clearly indicating different colour-luminosity relationships for different environments. On the other hand, unlike \citetalias{Kelsey2021}, differences in $\alpha$ are only potentially significant for local properties, being strongest for local $\mstellar$ ($\sim3\sigma$). There is a known correlation between $x_1$ and age \citep{Howell2009, Neill2009, Johansson2013,Childress2014, Wiseman2021}, which is often better probed by local tracers \citep{Rigault2020,Nicolas2021}. A different $\alpha$ in different local environments therefore indicates that the strength of the stretch-luminosity relationship could be dependent on the age of the local stellar population. On the other hand, if both $x_1$ and the luminosity step are correlated with age, then $\alpha$ can absorb some of the step in the fitting process \citep{Dixon2021,Rose2021,Wiseman2022}, and any difference here is a consequence of that degeneracy.

\begin{table*}
\begin{center}
\caption{Differences in best-fit $\alpha$ and $\beta$ when splitting the sample based on environmental properties.}
\begin{threeparttable}
\begin{tabular}{c c c c c c c} 
\hline
\multicolumn{1}{c}{Property} & \multicolumn{3}{c}{$\alpha$} & \multicolumn{3}{c}{$\beta$} \\&  < DP\tnote{b} & > DP & Difference ($\sigma$)\tnote{c} & < DP & > DP & Difference ($\sigma$)\\
\hline
Global Mass\tnote{a} & $ 0.16 \pm 0.02 $ & $ 0.18 \pm 0.01 $ & 0.9 & $ 3.50 \pm 0.13 $ & $ 2.93 \pm 0.09 $ & 3.6\\
Local Mass & $ 0.16 \pm 0.01 $ & $ 0.20 \pm 0.01 $ & 2.8 & $ 3.38 \pm 0.11 $ & $ 2.88 \pm 0.11 $ & 3.2\\
\hline
Global U-R & $ 0.17 \pm 0.01 $ & $ 0.19 \pm 0.01  $ & 1.4 & $ 3.29 \pm 0.11 $ & $ 2.92 \pm 0.10 $ & 2.5\\
Local U-R & $ 0.16 \pm 0.01 $ & $ 0.19 \pm 0.01 $ & 2.1 & $ 3.43 \pm 0.11 $ & $ 2.81 \pm 0.10 $ & 3.4\\
\hline 
\end{tabular}
\begin{tablenotes}
\item[a] Mass in $\log (\mstellar / M_\odot)$
\item[b] DP refers to the \lq{Division Point}\rq\ location of the environmental property step. For example, \lq{$<$DP}\rq\ indicates the lower mass or bluer environments. 
\item[c] Difference is the quadrature sum difference in $\alpha$ or $\beta$ between subsamples.
\end{tablenotes}
\end{threeparttable}
\label{table:env_split-alpha-beta}
\end{center}
\end{table*}

\section{Host environments and colour-dependent distance measurements}\label{colour}

\subsection{Splitting the sample based on colour} \label{csplit}

Motivated by our findings from \sref{refit}, and \citetalias{Kelsey2021}; \citetalias{BroutScolnic2021}, \citet{Popovic2021, Popovic2022} which suggest that the environmental \lq steps\rq\ in SN luminosity may be driven by underlying relationships between SN $c$ and galaxy properties, we split the SN Ia sample into two based on the SN colour ($c \leq 0$ and $c > 0$), and analyse the relations between $\Delta \mu$ and environmental property for each subsample. We examine these relationships for both global and local host galaxy properties. 

\begin{figure*}
\begin{center}
\includegraphics[width=.43\linewidth]{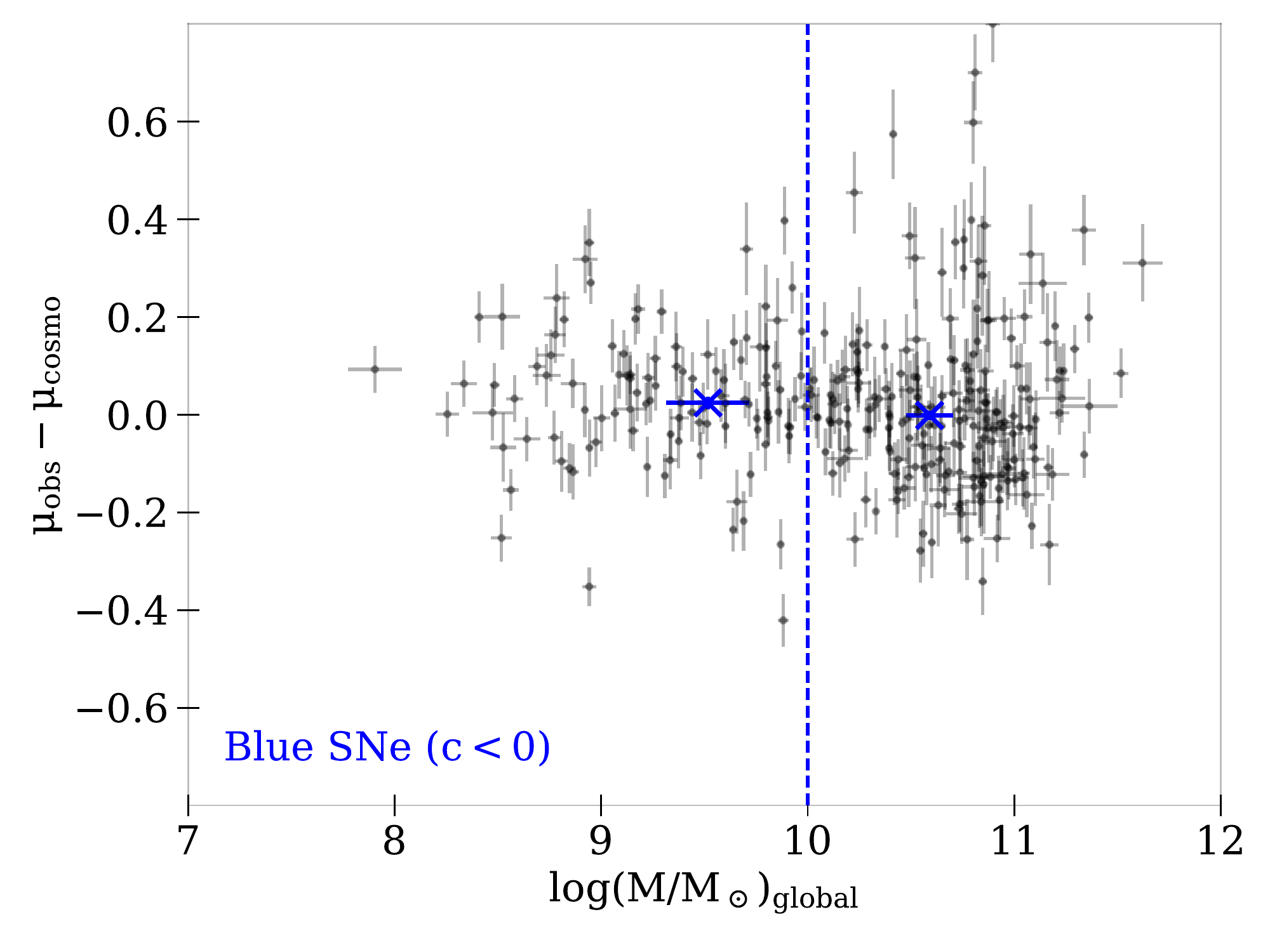}
\includegraphics[width=.43\linewidth]{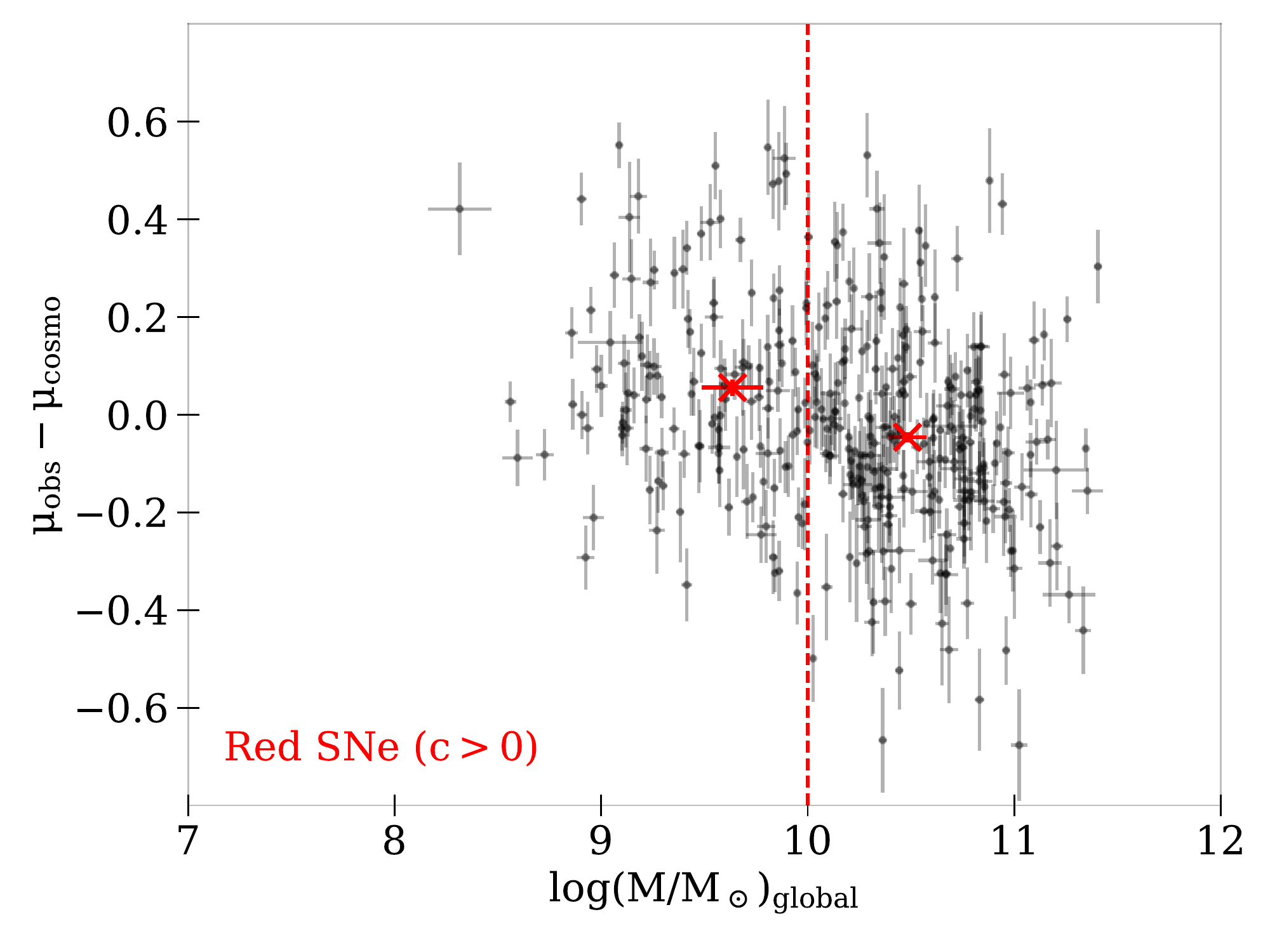}
\includegraphics[width=.43\linewidth]{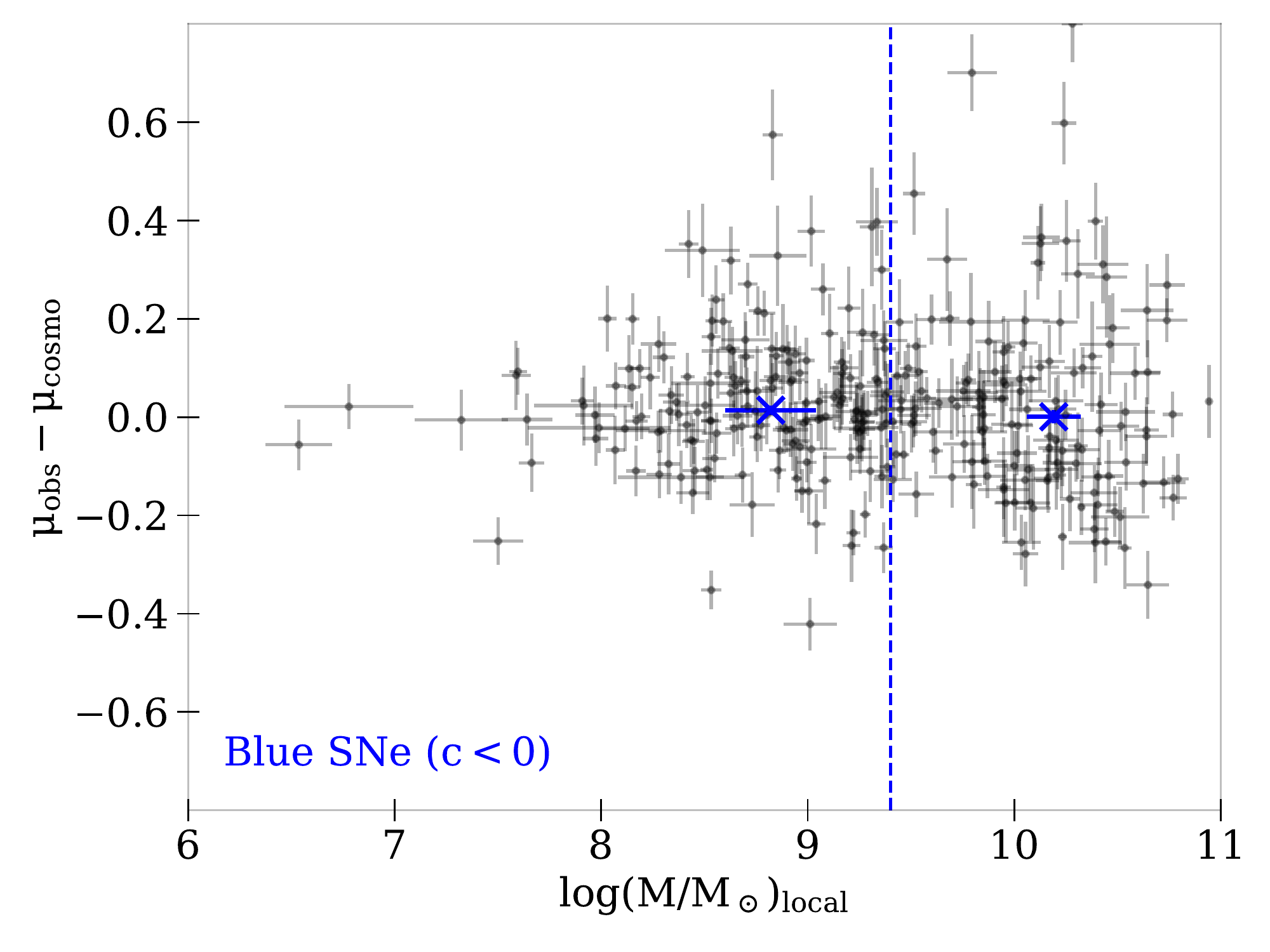}
\includegraphics[width=.43\linewidth]{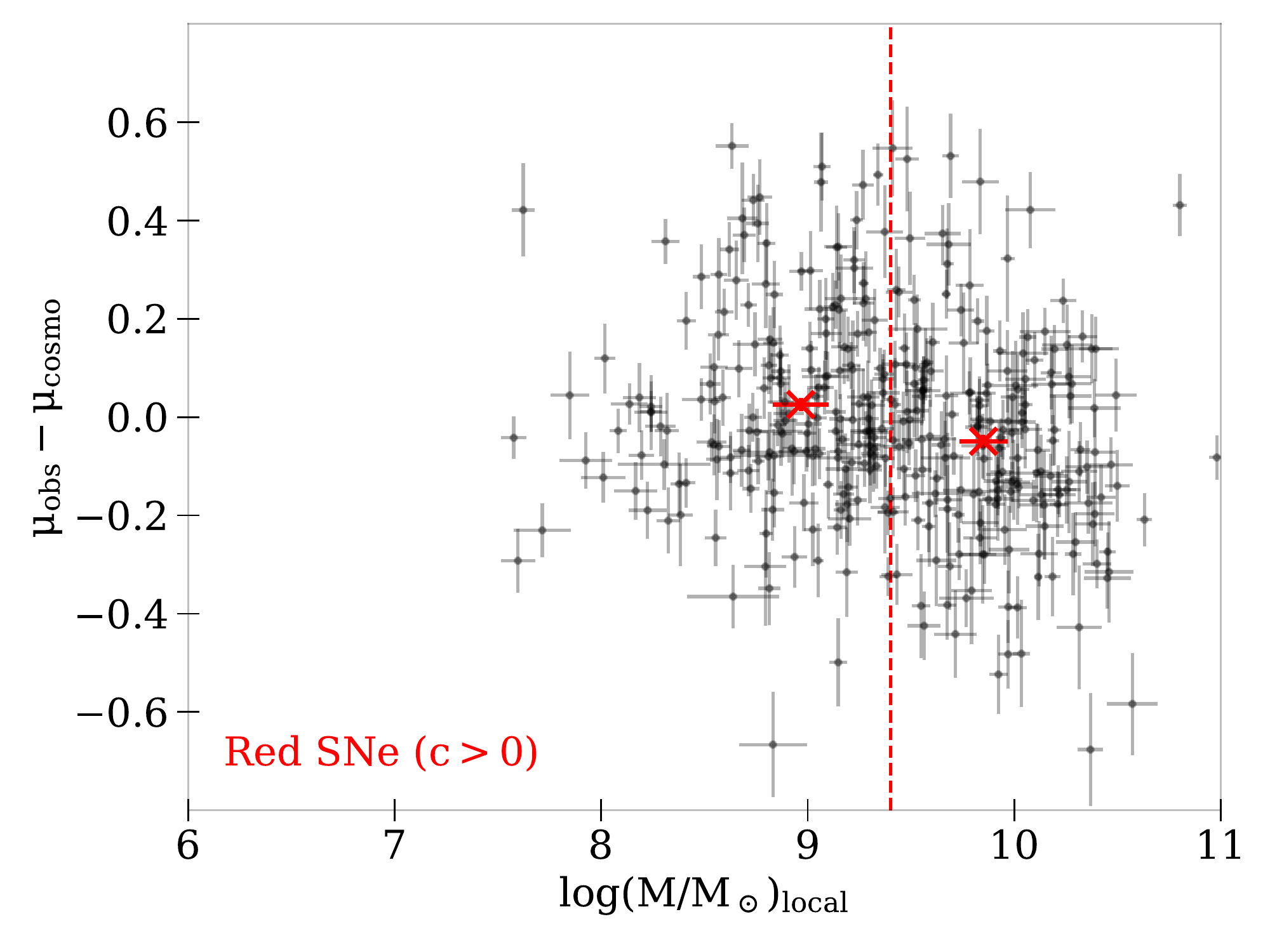}
\includegraphics[width=.43\linewidth]{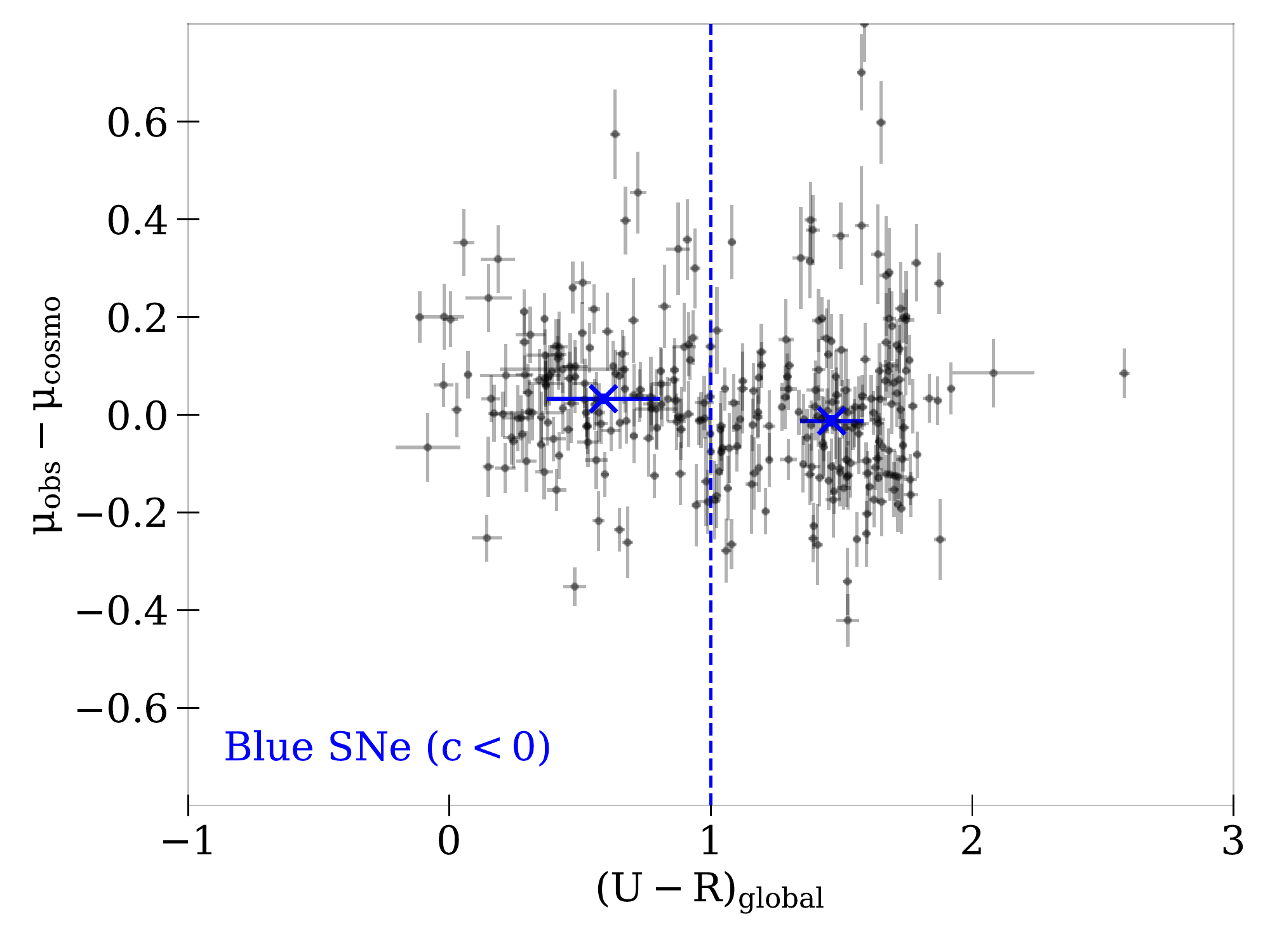}
\includegraphics[width=.43\linewidth]{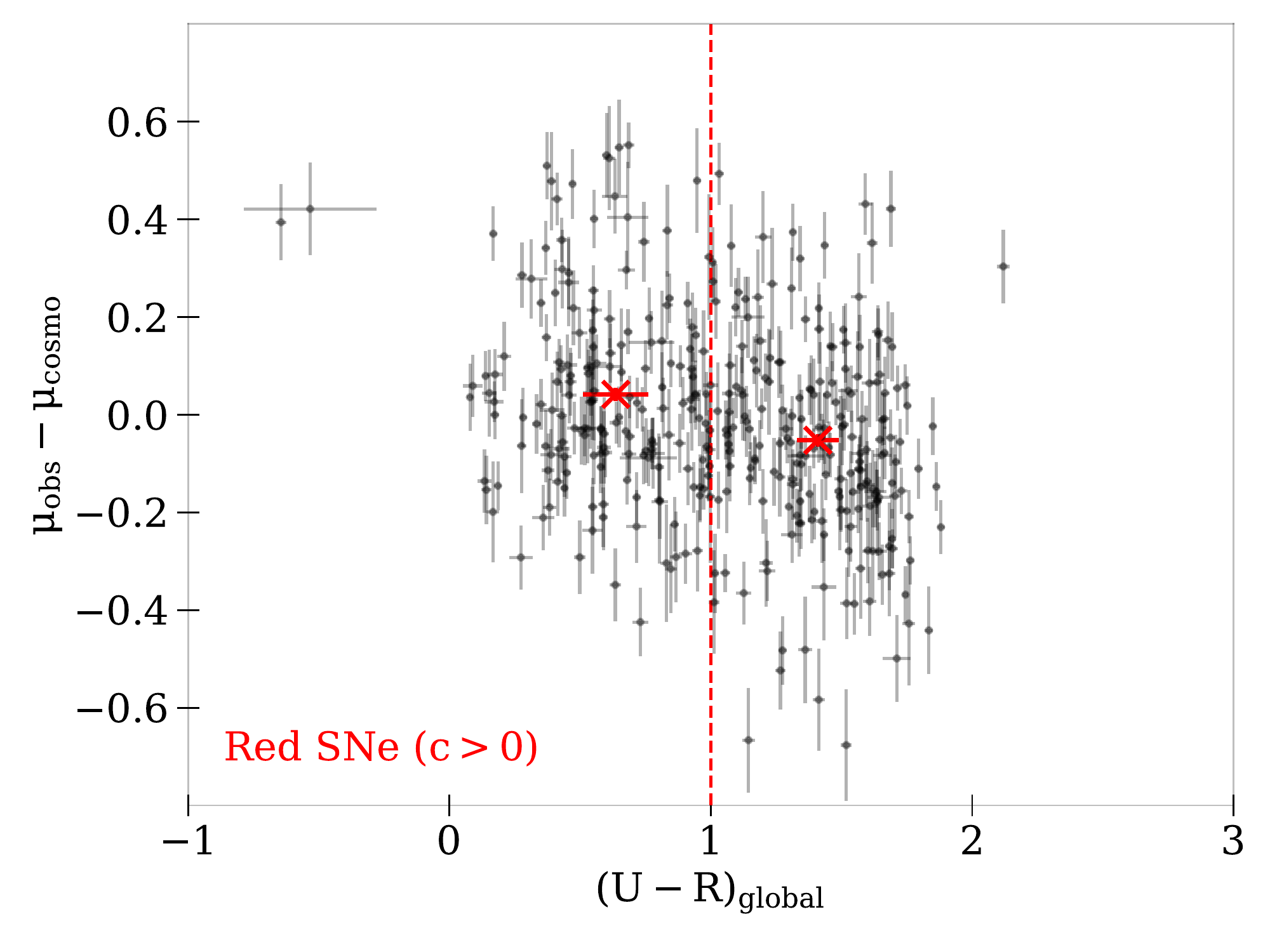}
\includegraphics[width=.43\linewidth]{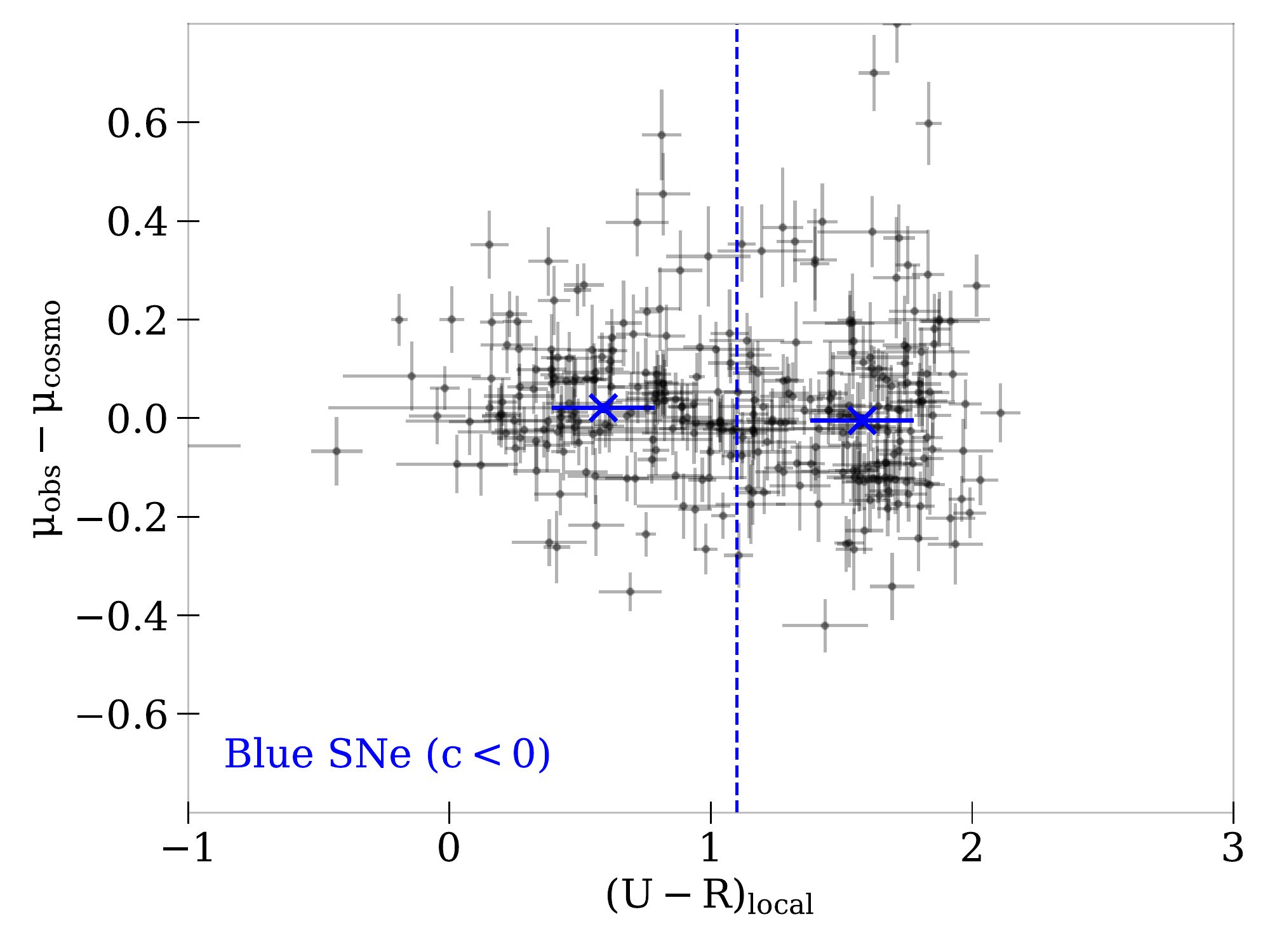}
\includegraphics[width=.43\linewidth]{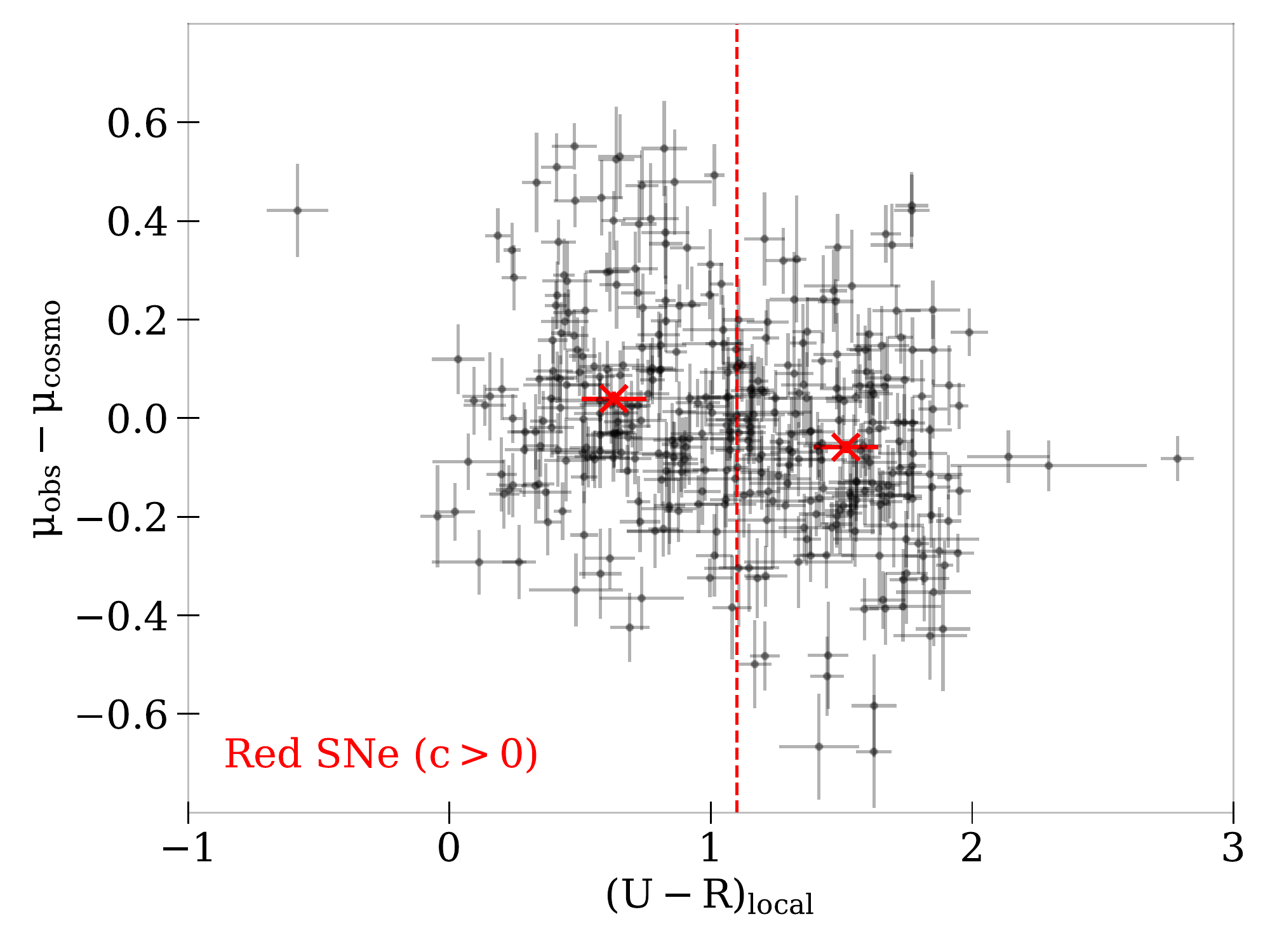}
\caption{Hubble residuals as a function of environmental properties for subsamples split by $c$. The dashed lines represent the split point for each sample indicated in \tref{table:c_split_steps_BBC1D}, defined as our location of the step, corresponding with the cross bin mean markers, displaying the steps in Hubble residual. These figures clearly show the differences in step size for red and blue SNe. Numerical values for these steps are displayed in \tref{table:c_split_steps_BBC1D} with r.m.s. values in \tref{table:c_split_rms_BBC1D}.}
\label{fig:csplit_steps}
\end{center}
\end{figure*}

The resulting steps for local and global $\mstellar$ and rest-frame $U-R$ for the different $c$ subsamples are displayed in \fref{fig:csplit_steps}, with numerical values given in \tref{table:c_split_steps_BBC1D}.

\begin{table*}
\begin{center}
\caption{Subsample Hubble residual steps when splitting the sample based on $c$ using a BBC1D correction.}
\begin{threeparttable}
\begin{tabular}{c c c c c c c} 
\hline
\multicolumn{2}{c}{Property} & \multicolumn{2}{c}{$c < 0$ Hubble Residual Step} & \multicolumn{2}{c}{$c > 0$ Hubble Residual Step} \\Name & Division Point & Magnitude & Sig. ($\sigma$)\tnote{b} & Magnitude & Sig. ($\sigma$) & Difference ($\sigma$)\tnote{c} \\
\hline
Number of Supernovae & & \multicolumn{2}{c}{306}& \multicolumn{2}{c}{369}&  \\
\hline
Global Mass\tnote{a} & 10.0 & $0.026\pm0.016$ & 1.6 & $0.102\pm0.020$ & 5.0 & 3.0\\
Local Mass & 9.4 & $0.013\pm0.016$ & 0.8 & $0.075\pm0.019$ & 4.0 & 2.5\\
\hline
Global U-R & 1.0 & $0.045\pm0.016$ & 2.9 & $0.094\pm0.019$ & 5.1 & 2.0\\
Local U-R & 1.1 & $0.026\pm0.016$ & 1.6 & $0.098\pm0.018$ & 5.3 & 3.0\\

\hline 
\end{tabular}
\begin{tablenotes}
\item[a] Mass in $\log (\mstellar / M_\odot)$
\item[b] Significance is quadrature sum.
\item[c] Difference is the quadrature sum difference in Hubble residual step magnitudes between red and blue subsamples.
\end{tablenotes}
\end{threeparttable}
\label{table:c_split_steps_BBC1D}
\end{center}
\end{table*}

In all cases, the step size is larger in red SNe Ia than in blue SNe Ia, but to varying levels of significance. There is a $3\sigma$ difference between Hubble residual step sizes for global $\mstellar$, as also seen in \citetalias{Kelsey2021}. This difference indicates that $\mstellar$ has a strong relationship with $c$, pointing to the link between host galaxy mass and dust. This is also consistent with the $3\sigma$ difference in step size for local $U-R$. The differences are not significant ($\sim2\sigma$) for the other environmental properties, indicating a weaker link between those properties and SN $c$. This result is different compared to \citetalias{Kelsey2021}, where all environmental property steps have differences of $>2\sigma$ when split into subsamples by $c$.

As in \citetalias{Kelsey2021} and \citetalias{BroutScolnic2021}, the r.m.s. values, presented in \tref{table:c_split_rms_BBC1D}, for SNe Ia with $c < 0$ are considerably smaller than those for SNe Ia with $c > 0$, with the smallest values found for $c < 0$ in low stellar mass or blue environments ($\sim0.14$). This lends more weight to the argument posed in \citet{Gonzalez-Gaitan2020} and \citetalias{Kelsey2021} that SNe Ia in the lower mass, higher star-forming, bluer regions are a more homogeneous sample that may be better standard candles. Our sample of blue SNe Ia in blue or low $\mstellar$ environments also have lower r.m.s. scatter than those obtained for a NIR sample \citep{Jones2022}, potentially raising questions about the necessity of space-based observations for SNe Ia cosmology. 

\begin{table*}
\begin{center}
\caption{Subsample Hubble residual r.m.s scatter when splitting the sample based on $c$ using a BBC1D correction.}
\begin{threeparttable}
\begin{tabular}{c c c c c c} 
\hline
\multicolumn{2}{c}{Property} & \multicolumn{2}{c}{$c < 0$ Hubble Residual r.m.s} & \multicolumn{2}{c}{$c > 0$ Hubble Residual r.m.s} \\Name & Division Point & $<$ DP\tnote{b} & $>$ DP & $<$ DP & $>$ DP\\
\hline
Global Mass\tnote{a} & 10.0 & 0.142 $\pm$ 0.019 & 0.174 $\pm$ 0.017 & 0.216 $\pm$ 0.027 & 0.208 $\pm$ 0.019\\
Local Mass & 9.4 & 0.142 $\pm$ 0.016 & 0.185 $\pm$ 0.022 & 0.204 $\pm$ 0.021 & 0.218 $\pm$ 0.023\\
\hline
Global U-R & 1.0 & 0.146 $\pm$ 0.018 & 0.176 $\pm$ 0.019 & 0.205 $\pm$ 0.022 & 0.216 $\pm$ 0.022\\
Local U-R & 1.1 & 0.141 $\pm$ 0.016 & 0.182 $\pm$ 0.020 & 0.210 $\pm$ 0.022 & 0.212 $\pm$ 0.022\\
\hline 
\end{tabular}
\begin{tablenotes}
\item[a] Mass in $\log (\mstellar / M_\odot)$
\item[b] DP refers to the \lq{Division Point}\rq\ location of the environmental property step. For example, \lq{$<$DP}\rq\ indicates the lower mass or bluer environments. 
\end{tablenotes}
\end{threeparttable}
\label{table:c_split_rms_BBC1D}
\end{center}
\end{table*}

\begin{figure*}
\begin{center}
\includegraphics[width=\linewidth]{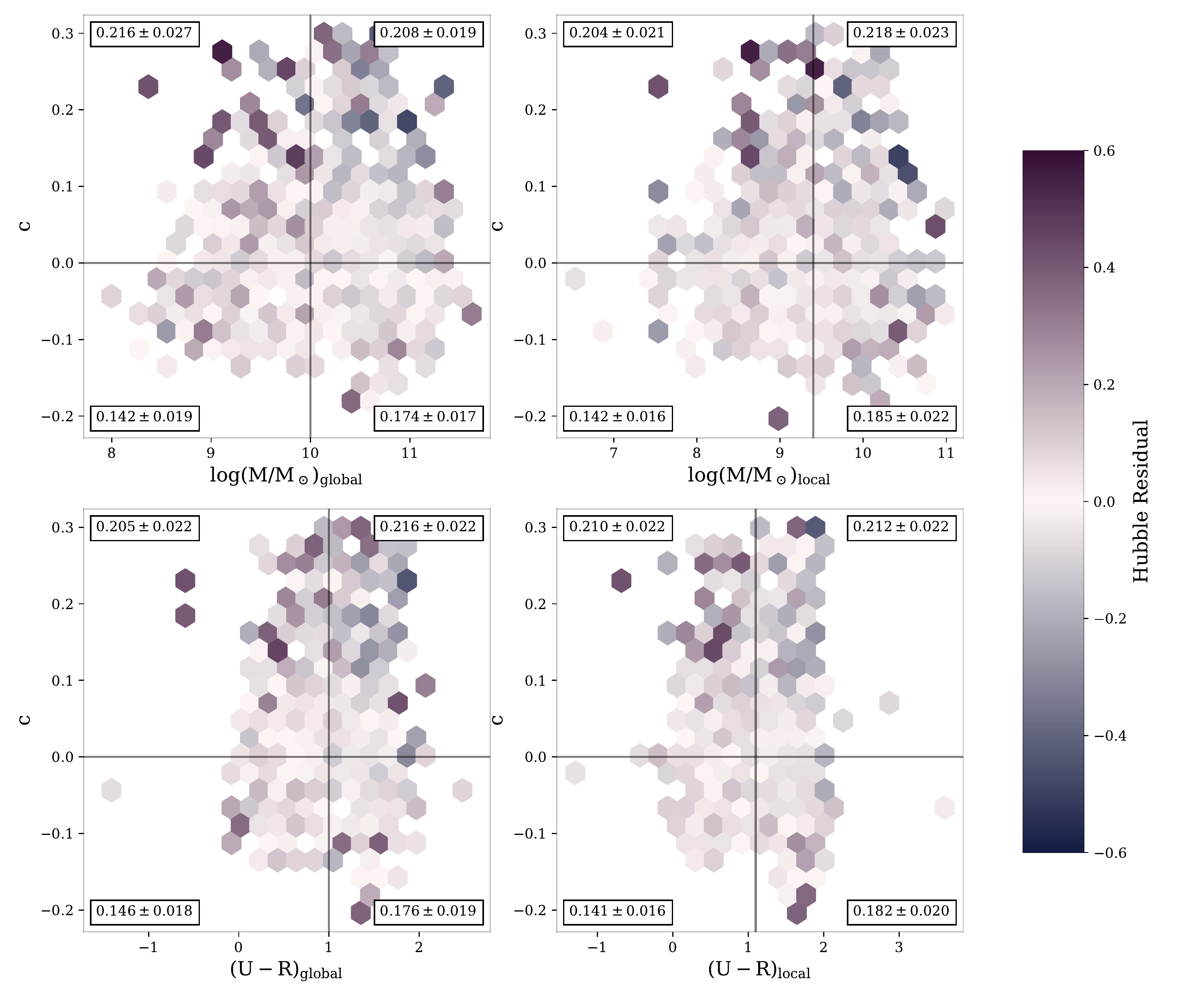}
\caption{Hexbinned heatmaps, showing the relationships between rest-frame $U-R$ or $\mstellar$ and $c$ as a function of mean Hubble residual. Vertical and horizontal lines show the splits into low and high environmental property, and blue ($c\leq0$) and red ($c>0$) colour SNe. The numbers in each quadrant are the r.m.s. values for the SN Ia Hubble residual scatter for events in that quadrant (also in \tref{table:c_split_rms_BBC1D}). Shown for both global (right panels) and local (left panels) $\mstellar$ (upper panels) and rest-frame $U-R$ galaxy colour (lower panels).}
\label{fig:quads}
\end{center}
\end{figure*}

The relationships with $c$ and Hubble residuals are presented in a different form in \fref{fig:quads}, using hexbinned heatmaps in the parameter space of environmental property and SN $c$, with bins shaded according to the mean Hubble residual of events in that bin. These plots show that the most homogeneous SN Ia sample with close to zero Hubble residual is in the lower left quadrants, indicating bluer SNe and low $\mstellar$ and/or blue $U-R$ regions.

\subsection{Comparison to \citet{BroutScolnic2021}} \label{BS20}

\citetalias{BroutScolnic2021} suggest that the dominant component of SN Ia intrinsic scatter is caused by variation in the total-to-selective extension ratio $R_{V}$ distribution as a function of host galaxy properties. They found that the Hubble residual trends with host $\mstellar$ were modelled well by considering the SNe $c$ distribution to be a two-component combination consisting of an intrinsic Gaussian distribution, and an extrinsic exponential $E(B-V)$ dust distribution. This extrinsic dust distribution is host galaxy $\mstellar$ dependent, with a Gaussian $R_V$ distribution, where mean $R_V = 2.75$ in low mass host galaxies and mean $R_V = 1.5$ in high mass hosts. The different $R_V$ values result in different effective colour-luminosity relationships either side of the mass step division point. \citetalias{BroutScolnic2021} suggest that the mass step is therefore primarily caused by a difference in dust properties for SNe Ia with different $c$. This interpretation is consistent with the finding in \citetalias{Kelsey2021} - it is physics that affects the SN colour that is driving the Hubble residual host galaxy correlations. 
To compare our analysis with \citetalias{BroutScolnic2021}, we extend the study of SN $c$ for different host properties, by comparing the Hubble residuals with a finer binning of SN colour, rather than simply red ($c>0$) or blue ($c<0$). This follows \citetalias{BroutScolnic2021} Fig.~6.

\subsubsection{Global $\mstellar$}\label{GlobM}

\begin{figure*}
\begin{center}
\begin{subfigure}[b]{0.49\textwidth}
    \centering
    \includegraphics[width=\linewidth]{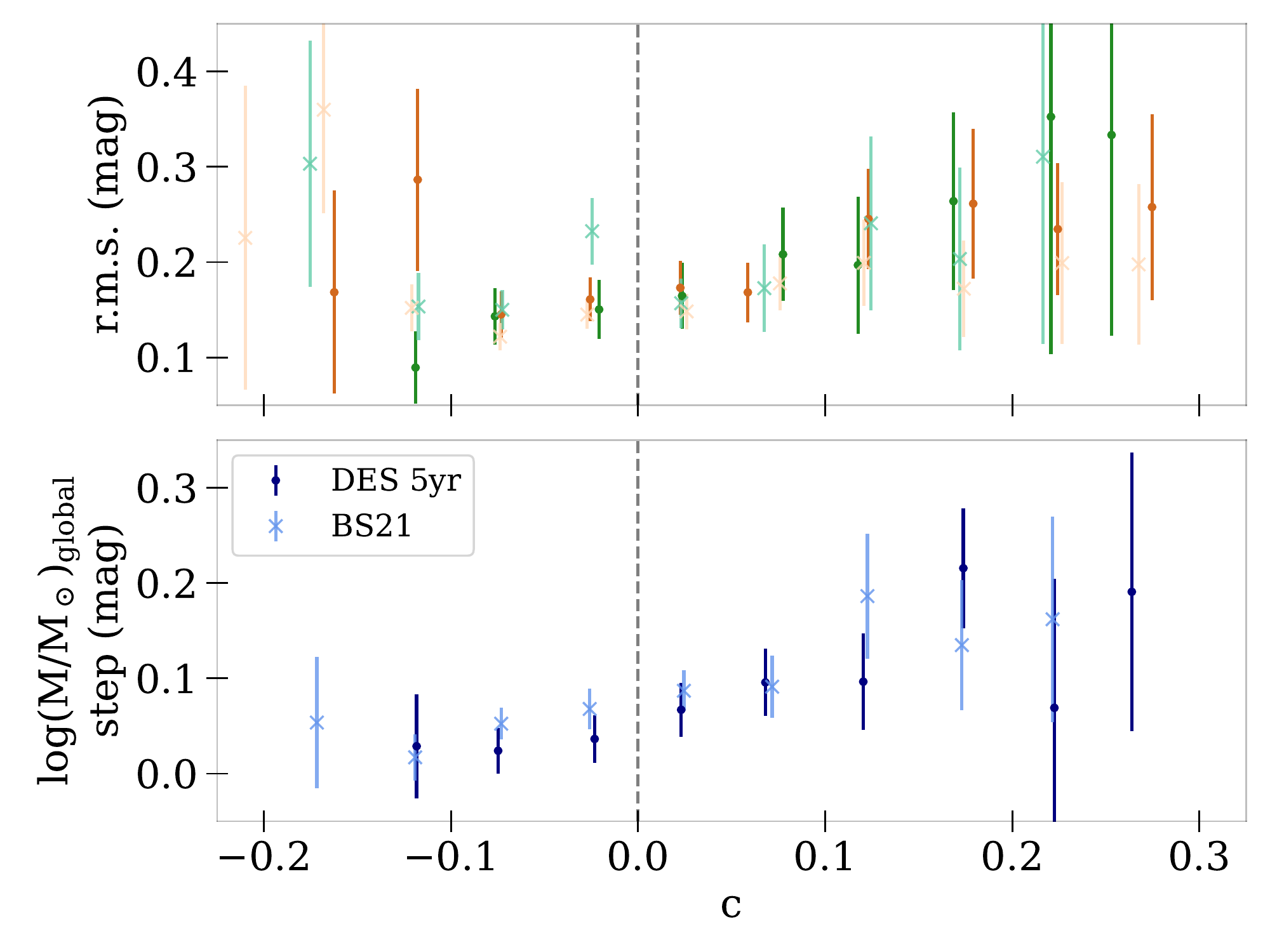}
    \caption{}
\end{subfigure}
\begin{subfigure}[b]{0.49\textwidth}
    \centering
    \includegraphics[width=\linewidth]{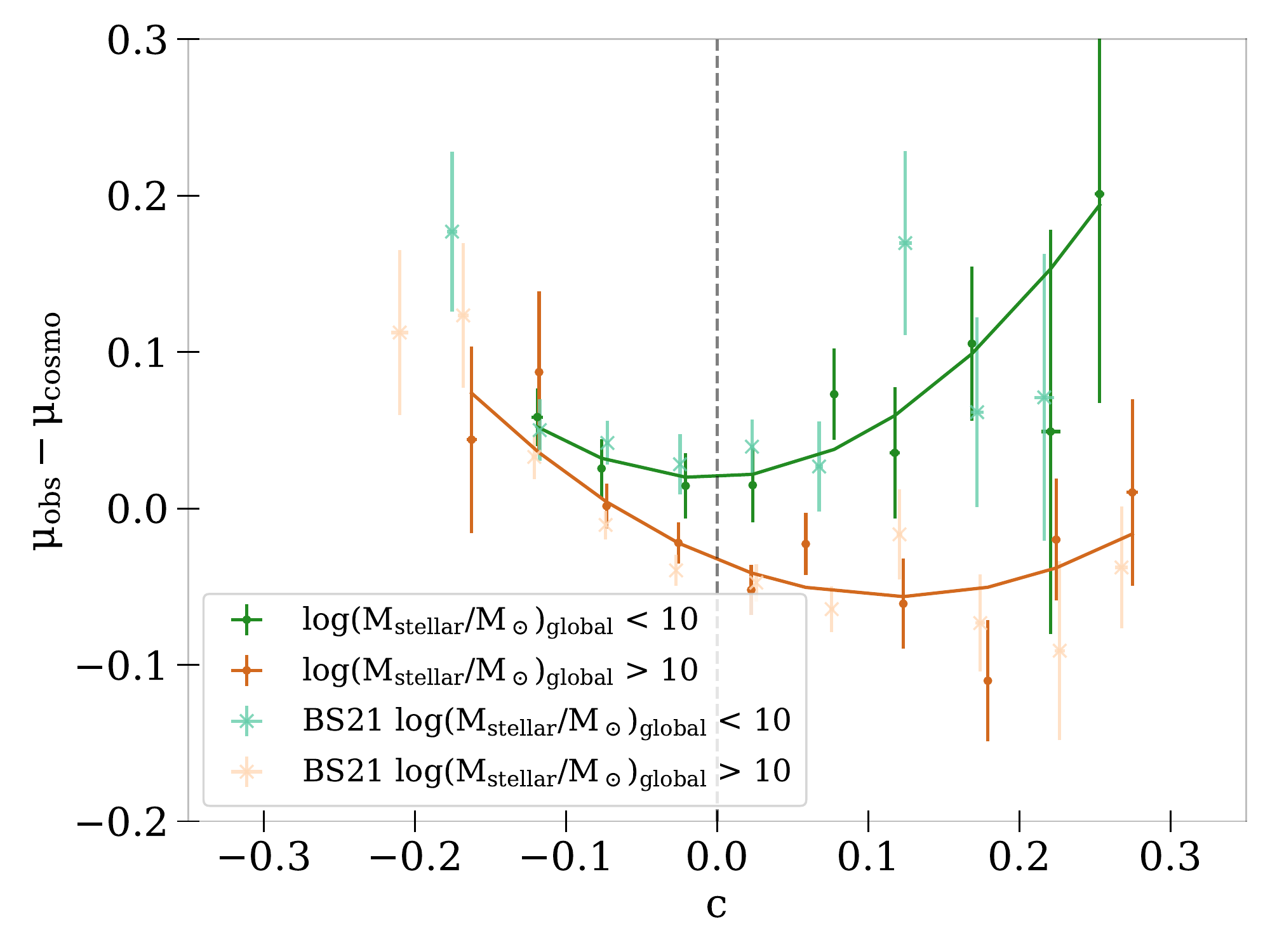}
    \caption{}
\end{subfigure}
\begin{subfigure}[b]{0.49\textwidth}
    \centering
    \includegraphics[width=\linewidth]{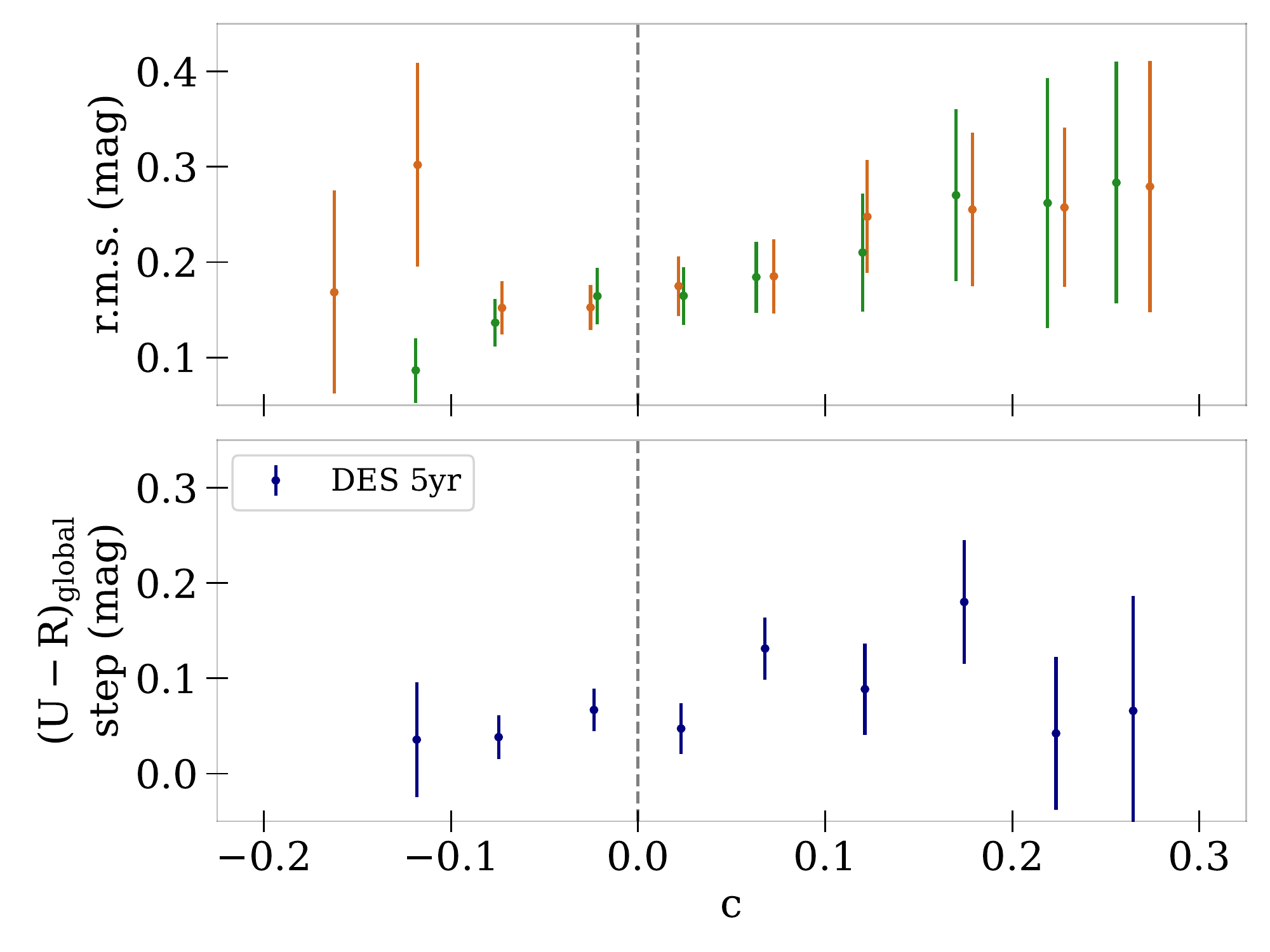}
    \caption{}
\end{subfigure}
\begin{subfigure}[b]{0.49\textwidth}
    \centering
    \includegraphics[width=\linewidth]{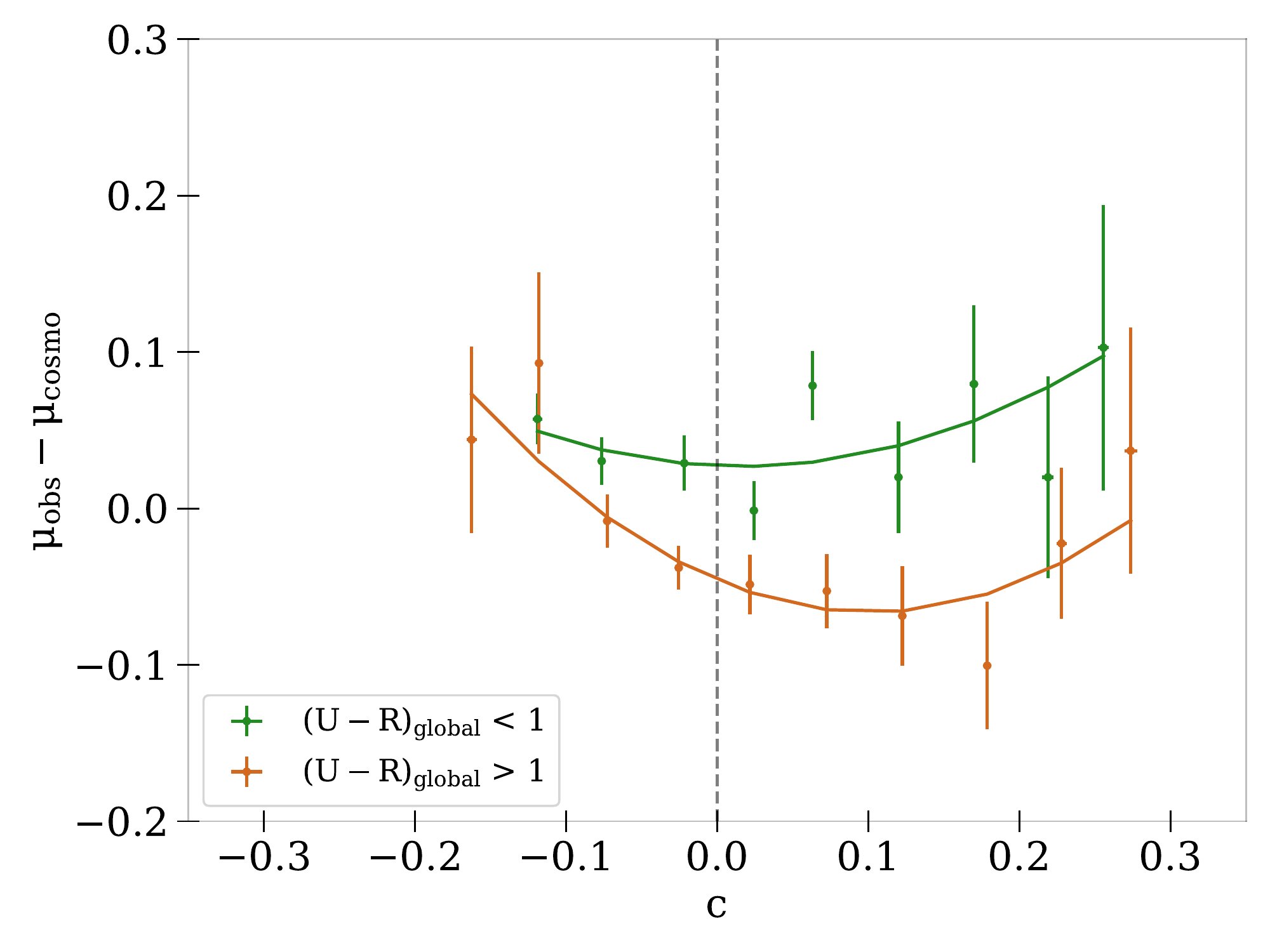}
    \caption{}
\end{subfigure}
\caption{\textbf{a)} and \textbf{c)} (Upper panels) Hubble diagram r.m.s. in bins of SALT2 colour $c$, for SNe Ia in galaxies with high and low $\mstellar$ a), or with high and low rest-frame $U-R$ colour c). Colour scheme corresponds with b) and d). (Lower panels) Calculated values for the size of the environmental property step as a function of $c$.  \textbf{b)} and \textbf{d)} Binned Hubble residuals as a function of $c$ split by host galaxy $\mstellar$ b) and host galaxy rest-frame $U-R$ d). The overplotted quadratic fits minimise the $\chi^2$. Data used in BS21 shown for $\mstellar$ in a) and b) in transparent colours. BS21 do not consider $U-R$, hence is not displayed in c) and d).}
\label{fig:likeBS20}
\end{center}
\end{figure*}

First, we present the results with host galaxy $\mstellar$ (\fref{fig:likeBS20}). Overplotted is the SN Ia sample used in \citetalias{BroutScolnic2021} (a mostly independent publicly available, spectroscopically classified, photometric light-curve sample consisting of a combination of data from the Foundation, PS1, SNLS, SDSS, CSP, CfA surveys\footnote{Foundation: \citet{Foley2018}, Pan-STARRS1 (PS1): \citet{Rest2014, Scolnic2018}, SuperNova Legacy Survey (SNLS): \citet{Betoule2014}, Sloan Digital Sky Survey (SDSS): \citet{Sako2011}, Carnegie Supernova Project (CSP): \citet{Stritzinger2010}, Harvard-Smithsonian Center for Astrophysics (CfA3+4): \citet{Hicken2009, Hicken2009b, Hicken2012}.}, and DES3YR) with a redshift cut of $z < 0.6$ applied for consistency with our analysis. The two data sets -- DES5YR and \citetalias{BroutScolnic2021} -- generally follow similar trends, and thus we expect that the predictions of the \citetalias{BroutScolnic2021} model will adequately model the relationships between environmental properties and $c$ for our DES5YR sample. \fref{fig:likeBS20}(a), as in \citetalias{BroutScolnic2021} Fig.~6, shows little difference between the r.m.s. values for samples in high and low $\mstellar$ for the bluer SNe, but this difference increases for the red SNe, also mirrored in the larger step sizes in the red bins. This increase in r.m.s. scatter and host $\mstellar$ step size towards the redder (right hand) end of the plot suggests that the overall $\mstellar$ step is driven by the red SNe. 

A polynomial is used to fit the Hubble residual versus $c$ relation for both high- and low-mass galaxies separately, minimising the $\chi^2$ in each case. These curves are then subtracted to remove the mass step from the data, enabling underlying environmental relationships to be uncovered. Similarly low $\chi^2$ values were found when fitting quadratic curves and when fitting two separate linear relations for positive and negative $c$, for both low and high $\mstellar$. These linear fits resulted in similar remaining relationships once their trends were removed from the data. However, we proceed with the quadratic fits, due to the fact that they are smooth, continuous functions. There is no clear reason why the colour-luminosity relation would change dramatically at any particular $c$ value, intuitively it is more likely to be a continuous relationship, meaning that combining linear functions for different $c$ bins may not be as realistic. As illustrated in \fref{fig:likeBS20}(b), these quadratic fits generate simple functions for the $\mstellar$ - $c$ dependent Hubble residual relationships. By subtracting these curves from the Hubble residual of each SN in our sample, we correct for these observed $c$ - dependent $\mstellar$ trends. As shown in \tref{table:fitremain_BBC1D}, this simple approximation of the \citetalias{BroutScolnic2021} dust model removes the global host galaxy mass step from our data ($0.001\pm0.013 \,\textrm{mag};\ 0.1\sigma$), however we find remaining rest-frame $U-R$ steps of $0.025\pm0.012 \,\textrm{mag};\ 2.1\sigma$ for global and $0.023\pm0.012 \,\textrm{mag};\ 1.9\sigma$ for local when the $c$ - dependent $\mstellar$ relation is removed, perhaps suggesting that the $\mstellar$ dust model is not the full picture, and should include or be fully based on the $U-R$ tracer instead (see \sref{GlobUR}). 

\begin{table*}
\begin{center}
\caption{Magnitudes and significances of remaining environmental property steps when fitting for relationships between $c$ and environmental properties using a BBC1D bias correction.}
\begin{threeparttable}
\begin{tabular}{c c c c c c} 
\hline
\multicolumn{6}{c}{Fitting For:}\\
 & & Host Mass & Local Mass & Host U-R & Local U-R \\
\hline
\parbox[t]{2mm}{\multirow{8}{*}{\rotatebox[origin=c]{90}{Remaining Step In:}}} & {\multirow{2}{*}{Host Mass}} & $0.001\pm$0.013 & $0.039\pm0.013$ & $0.012\pm0.013$ & $0.023\pm0.013$ \\
& & $0.1\sigma$ & $3.0\sigma$ & $1.0\sigma$ & $1.8\sigma$\\
& {\multirow{2}{*}{Local Mass}} & $0.011\pm0.012$ & $0.001\pm0.012$ & $0.010\pm0.012$ & $0.009\pm0.012$ \\
& & $0.9\sigma$ & $0.1\sigma$ & $0.8\sigma$ & $0.7\sigma$\\
& {\multirow{2}{*}{Host U-R}} & $0.025\pm0.012$ & $0.047\pm0.012$ & $0.001\pm0.012$ & $0.019\pm0.012$\\
& & $2.1\sigma$ & $3.8\sigma$ & $0.1\sigma$ & $1.6\sigma$\\
& {\multirow{2}{*}{Local U-R}} & $0.023\pm0.012$ & $0.037\pm0.012$ & $0.006\pm0.012$ & $0.001\pm0.012$\\
& & $1.9\sigma$ & $3.0\sigma$ & $0.5\sigma$ & $0.1\sigma$\\
\hline 
\end{tabular}
\begin{tablenotes}
\item All steps given in mag.
\item Step division points as described in Section~\ref{csplit}.
\end{tablenotes}
\end{threeparttable}
\label{table:fitremain_BBC1D}
\end{center}
\end{table*}

Whilst the remaining $U-R$ steps are small in our analysis, \citet{Roman2018} found a significant ($5\sigma$) remaining $U-V$ step when correcting first for the overall global mass step in their analysis (not dependent on $c$), suggesting that additional information can be provided by local properties when combined with global properties. Our results also agree with \citet{Galbany2022}, where  $>2\sigma$ steps in sSFR and H$\alpha$ equivalent width remain once the mass step has been corrected for, whilst a $<2\sigma$ step in $\mstellar$ remains once the reverse is done. This suggests that sSFH and H$\alpha$ equivalent width (both related to the age of the stellar population) are better than $\mstellar$ at improving SN Ia standardisation. 

\subsubsection{Global $U-R$} \label{GlobUR}

We repeated the above analysis, but starting with and fitting for the relationship between global rest-frame $U-R$ and $c$, instead of fitting for the global $\mstellar$ - $c$ dependent Hubble residual relationship. We split into \lq{low}\rq\ and \lq{high}\rq\ by splitting at $U-R = 1$, as motivated by \citetalias{Kelsey2021}. Again, quadratic functions fit the data best through a $\chi^2$ minimisation, which we subtracted from the Hubble residual for each SN to correct for $U-R$ - $c$ dependent Hubble residual relationships, as shown in panels (c) and (d) of \fref{fig:likeBS20}. 

As shown in \tref{table:fitremain_BBC1D}, this approximate correction removes the global $U-R$ step from the data, and we find remaining mass steps of only $0.012\pm0.013\,\textrm{mag} (1\sigma)$ for global $\mstellar$ and $0.010\pm0.012\,\textrm{mag} (0.8\sigma)$ for local. These post-correction steps are smaller than the remaining $U-R$ steps after the global $c$-$\mstellar$ relation was removed, suggesting that a $U-R$ correction encompasses more of the overall Hubble residual vs host environment relationship than the $\mstellar$ correction; as seen in \citetalias{Kelsey2021}. 

\subsubsection{Local corrections}

We repeat the corrections of \sref{GlobM} and \sref{GlobUR}, but for local properties instead of global, again presenting our results in \tref{table:fitremain_BBC1D}. 

The $\geq3\sigma$ steps that remain in both local and global $U-R$ and in global $\mstellar$ when fitting for local $\mstellar$ are particularly interesting. Considerable remaining steps remain once the trend with $c$ has been removed, suggesting that local mass may not be removing any trends or perhaps is less correlated with the other parameters than expected, however this disagrees with the trends shown in \citetalias{Kelsey2021} that local and global mass are correlated (albeit with scatter). This finding suggests that local mass may not be linked to dust in the same way as suggested by \citetalias{BroutScolnic2021} for global mass. We note in particular that local mass is the only parameter with a $<1\sigma$ step for $c<0$ (\tref{table:c_split_steps_BBC1D}), so may not follow the same trends with $c$ as the other parameters. Local mass can be understood as a stellar density, tracing the population of old stars in the region, so may be linked to age. Further investigation of this finding is needed in future study, and may require better resolved local properties than those available using DES. Higher resolution may help to determine the location of dust in the host galaxy, either contained in the local circumstellar region around a SNe, or more dispersed throughout the global host galaxy.

For local $U-R$, after fitting for the $c$ dependent relationship, steps of $<2\sigma$ remained in all other properties. This is likely reflective of the key result for host $U-R$, that a $U-R$ correction encompasses more of the dispersion than an $\mstellar$ correction. Within a 4kpc radius, local $U-R$ may not be truly \lq{local}\rq\ enough to see a clear difference compared to global $U-R$.

\section{Discussion} \label{discussion}

Similarly to \citetalias{Smith2020}, \citetalias{BroutScolnic2021} and \citetalias{Kelsey2021}, we find a $\sim3\sigma$ significant difference between global $\mstellar$ step sizes when splitting into subsamples based on SN $c$. The data agrees well with the dust explanation of \citetalias{BroutScolnic2021} and thus it is likely that the $\mstellar$ Hubble residual step differences for $c$ subsamples are due to the role of dust. 

However, with the larger sample afforded by the DES5YR photometric sample, we see a different result to \citetalias{Kelsey2021} with regards to global $U-R$ steps when splitting based on $c$. \citetalias{Kelsey2021} found $\sim3\sigma$ differences in step sizes in global $U-R$ between red and blue SNe, as opposed to the smaller $\sim2\sigma$ difference seen with this DES5YR sample. 

As the observed trend with colour is consistent with the \citetalias{BroutScolnic2021} dust model, we can remove the effect of the mass step from the data by fitting for such a $c$-dependent global host $\mstellar$ Hubble residual relationship, and subtracting it from the Hubble residuals. Such a method has been introduced as a \lq{4D}\rq\ bias correction in \citet{Popovic2021}, and has been shown to result in a $\sigma w_{\textrm{sys}} \sim 0.005$ \citep{Popovic2022}. However, in this analysis we found an intriguing $2\sigma$ remaining global and local $U-R$ steps once the mass step has been removed, indicating that, whilst $\mstellar$-based dust modelling may explain the mass step \citepalias{BroutScolnic2021}, it may not fully explain the SN luminosity dispersion. Further investigation of this tentative result is needed.

Despite $\mstellar$ and $U-R$ being highly correlated, our analysis shows that the most Hubble residual dispersion across environmental properties was removed when correcting for a $c$-dependent global $U-R$ relation. As $U-R$ is connected to stellar age, this is expected given an older stellar population is one in which a larger fraction of the hotter stars have had time to explode and create dust. This result motivates further work into integrating mass and age simultaneously into scatter models and bias correction, with initial investigations presented in \citet{Wiseman2022}. 

\subsection{Impact on Cosmology}

Based on this analysis, our suggestion for future cosmology analyses is to correct for a global $c$-dependent $U-R$ effect as this step is largest and removes the most remaining dispersion of the local and global $U-R$ and $\mstellar$ environmental properties measured. Alternatively, corrections could be combined to remove more dispersion than one correction alone. To reduce potential bias in the standardisation, these should be simultaneously fit with the other light-curve standardisation parameters \citep{Rose2021, Dixon2021}.

However, given the homogeneity of blue ($c < 0$) supernovae in low mass or locally blue environments (as shown in \tref{table:c_split_rms_BBC1D} and \fref{fig:quads}), it may be simplest and of most immediate value to use these SNe in cosmology \citep{Kelsey2021, Gonzalez-Gaitan2020}, mitigating the need for environment correction. This is not a new suggestion, and there is a wealth of information pointing to the benefits of such a cut. For example, \citet{Rigault2013} postulate that SNe Ia from locally passive environments are the cause of the biases they observed due to their higher scatter, and they suggest adding a selection cut to only include those in locally star forming (i.e. blue) environments for cosmology. This is emphasised by  \citet{Childress2014}, \citet{Kelly2015}, \citet{Henne2017} and \citet{Kim2018} who all find consistent results, and make the same conclusions about selecting star forming galaxies. \citet{Graur2015} and \citet{Kim2018} both suggest that the scatter is further constrained by limiting to low-mass ($\leq 10^{10} M_\odot$) globally star-forming host galaxies. In another test, through the analysis of ejecta velocities, \citet{Wang2009}, \citet{FoleyKasen2011} and \citet{Siebert2020} all find that the SN scatter can be reduced by using lower-velocity, bluer supernovae. By combining all of this knowledge from previous analyses, and the confirmations from \citetalias{Kelsey2021}, \citet{Gonzalez-Gaitan2020} and this study, we should use a subset of blue ($c < 0$) SNe in low mass/blue/star-forming environments to provide the most homogeneous sample for future cosmology. 

Reducing the size of a SN Ia sample by 75 per cent to obtain a sample of blue events in blue environments would roughly double the size of the statistical uncertainties, all things being equal. However, this loss of precision is offset by both the 30 per cent reduction in Hubble residual scatter for these SNe Ia, as well as additional reductions in systematic uncertainties, such as removing the need to correct for the luminosity step and its associated astrophysical uncertainties. In an analysis such as the Time Domain Extragalactic Survey \citep[TiDES;][]{Swann2019} where the sample size will be tens of thousands of SNe Ia, the statistical uncertainty will be below the level of the systematic uncertainty (Frohmaier et al. in prep), so such a selection to reduce scatter and other systematics offers clear advantages for cosmology.

\section{Summary} \label{summary}

By expanding the findings of our previous study into the relationship between SNIa host environment and $c$ \citepalias{Kelsey2021} to a larger sample consisting of SNIa from DES5YR, we have provided more weight to suggestions for future cosmological analyses, and have added our point of view to the historic mass vs age debate.

From our analysis our key findings are as follows:
\begin{enumerate}
    \item Hubble residual steps in environmental properties are consistent with prior analyses, with values of $\sim5\sigma$ for global $\mstellar$, and for global and local rest-frame $U-R$. The local mass step is slightly smaller at $\sim4\sigma$.
    \item When splitting our data into subsamples based on $c$, the largest, and most statistically significant, differences in Hubble residual \lq{step}\rq\ are associated with host \mstellar\ and local $U-R$, agreeing with \citetalias{Kelsey2021}. 
    \item As in \citetalias{Kelsey2021} and \citet{Gonzalez-Gaitan2020}, we observe the lowest rms scatter, and thus highest homogeneity for blue ($c < 0$) supernovae in low mass or blue environments. This suggests that such a subsample of supernovae may provide the best sample for use in future cosmological analyses. 
    \item Despite removing the mass step, intriguing $2\sigma$ steps in global and local $U-R$ remain after fitting for a simple approximation of the \citetalias{BroutScolnic2021} dust model. This suggests that current dust modelling may not fully explain the dispersion in SN luminosity.
    \item The remaining dispersion is minimised considering a $c$-dependent global $U-R$ relation (i.e. leaves the lowest-significance residual steps in the other parameters), implying that $U-R$ provides different information about the environment of SNe Ia than $\mstellar$.
\end{enumerate}

This analysis has important cosmological implications, which should be taken into account in the next generation of cosmological analyses. On one hand, the homogeneity of blue SNe in low mass or blue environments provides more weight to the argument that they are the best subsample to use for precision cosmology, so it may simply be easiest to just use those. On the other hand, to gain insight into the true astrophysical cause of the SNe Ia dispersion, combining environmental corrections or studying the impact of dust on galaxy $U-R$ may provide the answers for the true relationships between SNe Ia and their environments. 

\section*{Acknowledgements}

This work was supported by the Science and Technology Facilities Council [grant number ST/P006760/1] through the DISCnet Centre for Doctoral Training. L.K. thanks the UKRI Future Leaders Fellowship for support through the grant MR/T01881X/1. M.S. acknowledges from EU/FP7-ERC grant 615929, and P.W. acknowledges support from STFC grant ST/R000506/1. L.G. acknowledges financial support from the Spanish Ministerio de Ciencia e Innovaci\'on (MCIN), the Agencia Estatal de Investigaci\'on (AEI) 10.13039/501100011033, and the European Social Fund (ESF) "Investing in your future" under the 2019 Ram\'on y Cajal program RYC2019-027683-I and the PID2020-115253GA-I00 HOSTFLOWS project, from Centro Superior de Investigaciones Cient\'ificas (CSIC) under the PIE project 20215AT016, and the program Unidad de Excelencia Mar\'ia de Maeztu CEX2020-001058-M.

This research made use of Astropy,\footnote{\url{http://www.astropy.org}} a community-developed core Python package for Astronomy \citep{astropy:2013, astropy:2018}. This research made use of Photutils, an Astropy package for detection and photometry of astronomical sources \citep{Bradley2019}.

Funding for the DES Projects has been provided by the U.S. Department of Energy, the U.S. National Science Foundation, the Ministry of Science and Education of Spain, 
the Science and Technology Facilities Council of the United Kingdom, the Higher Education Funding Council for England, the National Center for Supercomputing 
Applications at the University of Illinois at Urbana-Champaign, the Kavli Institute of Cosmological Physics at the University of Chicago, 
the Center for Cosmology and Astro-Particle Physics at the Ohio State University,
the Mitchell Institute for Fundamental Physics and Astronomy at Texas A\&M University, Financiadora de Estudos e Projetos, 
Funda{\c c}{\~a}o Carlos Chagas Filho de Amparo {\`a} Pesquisa do Estado do Rio de Janeiro, Conselho Nacional de Desenvolvimento Cient{\'i}fico e Tecnol{\'o}gico and 
the Minist{\'e}rio da Ci{\^e}ncia, Tecnologia e Inova{\c c}{\~a}o, the Deutsche Forschungsgemeinschaft and the Collaborating Institutions in the Dark Energy Survey. 

The Collaborating Institutions are Argonne National Laboratory, the University of California at Santa Cruz, the University of Cambridge, Centro de Investigaciones Energ{\'e}ticas, 
Medioambientales y Tecnol{\'o}gicas-Madrid, the University of Chicago, University College London, the DES-Brazil Consortium, the University of Edinburgh, 
the Eidgen{\"o}ssische Technische Hochschule (ETH) Z{\"u}rich, 
Fermi National Accelerator Laboratory, the University of Illinois at Urbana-Champaign, the Institut de Ci{\`e}ncies de l'Espai (IEEC/CSIC), 
the Institut de F{\'i}sica d'Altes Energies, Lawrence Berkeley National Laboratory, the Ludwig-Maximilians Universit{\"a}t M{\"u}nchen and the associated Excellence Cluster Universe, 
the University of Michigan, the National Optical Astronomy Observatory, the University of Nottingham, The Ohio State University, the University of Pennsylvania, the University of Portsmouth, 
SLAC National Accelerator Laboratory, Stanford University, the University of Sussex, Texas A\&M University, and the OzDES Membership Consortium.

Based in part on observations at Cerro Tololo Inter-American Observatory, National Optical Astronomy Observatory, which is operated by the Association of 
Universities for Research in Astronomy (AURA) under a cooperative agreement with the National Science Foundation.

The DES data management system is supported by the National Science Foundation under Grant Numbers AST-1138766 and AST-1536171.
The DES participants from Spanish institutions are partially supported by MINECO under grants AYA2015-71825, ESP2015-66861, FPA2015-68048, SEV-2016-0588, SEV-2016-0597, and MDM-2015-0509, 
some of which include ERDF funds from the European Union. IFAE is partially funded by the CERCA program of the Generalitat de Catalunya.
Research leading to these results has received funding from the European Research
Council under the European Union's Seventh Framework Program (FP7/2007-2013) including ERC grant agreements 240672, 291329, and 306478.
We  acknowledge support from the Brazilian Instituto Nacional de Ci\^encia
e Tecnologia (INCT) e-Universe (CNPq grant 465376/2014-2).

This manuscript has been authored by Fermi Research Alliance, LLC under Contract No. DE-AC02-07CH11359 with the U.S. Department of Energy, Office of Science, Office of High Energy Physics.
\section*{Data Availability}

The global and local photometry and derived environmental properties for the 675 SNe presented in this analysis are available in the online supplementary material. The light curves for the full DES-SN photometric SN Ia catalogue and associated host galaxy data will be made available as part of the DES5YR SN cosmology analysis at https://des.ncsa.illinois.edu/releases/sn.


\bibliographystyle{mnras}
\bibliography{biblio} 


\vspace{2cm}
\noindent
$^{1}$ Institute of Cosmology and Gravitation, University of Portsmouth, Portsmouth, PO1 3FX, UK\\
$^{2}$ School of Physics and Astronomy, University of Southampton,  Southampton, SO17 1BJ, UK\\
$^{3}$ The Research School of Astronomy and Astrophysics, Australian National University, ACT 2601, Australia\\
$^{4}$ Department of Physics, Duke University Durham, NC 27708, USA\\
$^{5}$ Einstein Fellow\\
$^{6}$ Center for Astrophysics $\vert$ Harvard \& Smithsonian, 60 Garden Street, Cambridge, MA 02138, USA\\
$^{7}$ School of Mathematics and Physics, University of Queensland,  Brisbane, QLD 4072, Australia\\
$^{8}$ Centre for Astrophysics \& Supercomputing, Swinburne University of Technology, Victoria 3122, Australia\\
$^{9}$ Institut d'Estudis Espacials de Catalunya (IEEC), 08034 Barcelona, Spain\\
$^{10}$ Institute of Space Sciences (ICE, CSIC),  Campus UAB, Carrer de Can Magrans, s/n,  08193 Barcelona, Spain\\
$^{11}$ Department of Astronomy and Astrophysics, University of Chicago, Chicago, IL 60637, USA\\
$^{12}$ Kavli Institute for Cosmological Physics, University of Chicago, Chicago, IL 60637, USA\\
$^{13}$ Centre for Gravitational Astrophysics, College of Science, The Australian National University, ACT 2601, Australia\\
$^{14}$ Univ Lyon, Univ Claude Bernard Lyon 1, CNRS, IP2I Lyon / IN2P3, IMR 5822, F-69622 Villeurbanne, France\\
$^{15}$ Cerro Tololo Inter-American Observatory, NSF's National Optical-Infrared Astronomy Research Laboratory, Casilla 603, La Serena, Chile\\
$^{16}$ Laborat\'orio Interinstitucional de e-Astronomia - LIneA, Rua Gal. Jos\'e Cristino 77, Rio de Janeiro, RJ - 20921-400, Brazil\\
$^{17}$ Fermi National Accelerator Laboratory, P. O. Box 500, Batavia, IL 60510, USA\\
$^{18}$ Department of Physics, University of Michigan, Ann Arbor, MI 48109, USA\\
$^{19}$ CNRS, UMR 7095, Institut d'Astrophysique de Paris, F-75014, Paris, France\\
$^{20}$ Sorbonne Universit\'es, UPMC Univ Paris 06, UMR 7095, Institut d'Astrophysique de Paris, F-75014, Paris, France\\
$^{21}$ University Observatory, Faculty of Physics, Ludwig-Maximilians-Universit\"at, Scheinerstr. 1, 81679 Munich, Germany\\
$^{22}$ Department of Physics \& Astronomy, University College London, Gower Street, London, WC1E 6BT, UK\\
$^{23}$ Kavli Institute for Particle Astrophysics \& Cosmology, P. O. Box 2450, Stanford University, Stanford, CA 94305, USA\\
$^{24}$ SLAC National Accelerator Laboratory, Menlo Park, CA 94025, USA\\
$^{25}$ Instituto de Astrofisica de Canarias, E-38205 La Laguna, Tenerife, Spain\\
$^{26}$ Universidad de La Laguna, Dpto. Astrofísica, E-38206 La Laguna, Tenerife, Spain\\
$^{27}$ Center for Astrophysical Surveys, National Center for Supercomputing Applications, 1205 West Clark St., Urbana, IL 61801, USA\\
$^{28}$ Department of Astronomy, University of Illinois at Urbana-Champaign, 1002 W. Green Street, Urbana, IL 61801, USA\\
$^{29}$ Institut de F\'{\i}sica d'Altes Energies (IFAE), The Barcelona Institute of Science and Technology, Campus UAB, 08193 Bellaterra (Barcelona) Spain\\
$^{30}$ Astronomy Unit, Department of Physics, University of Trieste, via Tiepolo 11, I-34131 Trieste, Italy\\
$^{31}$ INAF-Osservatorio Astronomico di Trieste, via G. B. Tiepolo 11, I-34143 Trieste, Italy\\
$^{32}$ Institute for Fundamental Physics of the Universe, Via Beirut 2, 34014 Trieste, Italy\\
$^{33}$ Hamburger Sternwarte, Universit\"{a}t Hamburg, Gojenbergsweg 112, 21029 Hamburg, Germany\\
$^{34}$ Department of Physics, IIT Hyderabad, Kandi, Telangana 502285, India\\
$^{35}$ Jet Propulsion Laboratory, California Institute of Technology, 4800 Oak Grove Dr., Pasadena, CA 91109, USA\\
$^{36}$ Institute of Theoretical Astrophysics, University of Oslo. P.O. Box 1029 Blindern, NO-0315 Oslo, Norway\\
$^{37}$ Instituto de Fisica Teorica UAM/CSIC, Universidad Autonoma de Madrid, 28049 Madrid, Spain\\
$^{38}$ Observat\'orio Nacional, Rua Gal. Jos\'e Cristino 77, Rio de Janeiro, RJ - 20921-400, Brazil\\
$^{39}$ Santa Cruz Institute for Particle Physics, Santa Cruz, CA 95064, USA\\
$^{40}$ Center for Cosmology and Astro-Particle Physics, The Ohio State University, Columbus, OH 43210, USA\\
$^{41}$ Department of Physics, The Ohio State University, Columbus, OH 43210, USA\\
$^{42}$ Australian Astronomical Optics, Macquarie University, North Ryde, NSW 2113, Australia\\
$^{43}$ Lowell Observatory, 1400 Mars Hill Rd, Flagstaff, AZ 86001, USA\\
$^{44}$ Sydney Institute for Astronomy, School of Physics, A28, The University of Sydney, NSW 2006, Australia\\
$^{45}$ Centro de Investigaciones Energ\'eticas, Medioambientales y Tecnol\'ogicas (CIEMAT), Madrid, Spain\\
$^{46}$ Instituci\'o Catalana de Recerca i Estudis Avan\c{c}ats, E-08010 Barcelona, Spain\\
$^{47}$ Department of Astronomy, University of California, Berkeley,  501 Campbell Hall, Berkeley, CA 94720, USA\\
$^{48}$ Institute of Astronomy, University of Cambridge, Madingley Road, Cambridge CB3 0HA, UK\\
$^{49}$ Department of Astrophysical Sciences, Princeton University, Peyton Hall, Princeton, NJ 08544, USA\\
$^{50}$ Department of Physics and Astronomy, University of Pennsylvania, Philadelphia, PA 19104, USA\\
$^{51}$ Department of Physics and Astronomy, Pevensey Building, University of Sussex, Brighton, BN1 9QH, UK\\
$^{52}$ Computer Science and Mathematics Division, Oak Ridge National Laboratory, Oak Ridge, TN 37831\\
$^{53}$ National Center for Supercomputing Applications, 1205 West Clark St., Urbana, IL 61801, USA\\
$^{54}$ Lawrence Berkeley National Laboratory, 1 Cyclotron Road, Berkeley, CA 94720, USA\\


\appendix 
\clearpage
\onecolumn
\section{BBC5D} \label{BBC5D}

In this analysis, we have focused on using a BBC1D bias correction, however for completeness and consistency with \citetalias{Kelsey2021}, here we discuss the differences in results when using a BBC5D bias correction. When using BBC5D, three fewer SNe Ia pass our quality cuts, resulting in a sample of 672.  Using SALT2 (Section~\ref{params}), we obtain values of $\alpha =  0.164\pm0.009 $ and $\beta =  3.36\pm0.07$ for this sample. 

For the overall environmental property steps, as presented in \tref{table:5yr_4values_BBC5D}, there is little difference in magnitude or significance, with only a slight decrease ($\sim0.01$\ mag) for BBC5D. The Hubble residual r.m.s. values decrease for all properties and for each side of the division point, but more so on the right hand side (higher $\mstellar$ or redder). This results in smaller differences in the r.m.s. values either side of the steps for BBC5D than for BBC1D, indicating that the additional corrections of BBC5D are absorbing some of the dispersion. 

\begin{table*}
\begin{center}
\caption{Hubble residual steps for stellar mass and $U-R$ for the DES5YR data using a 5D bias correction.}
\begin{threeparttable}
\begin{tabular}{c c c c c c}

\hline
\multicolumn{2}{c}{Property} &  \multicolumn{2}{c}{Hubble Residual Step} & \multicolumn{2}{c}{Hubble Residual r.m.s.}\\ Name & Division Point & Magnitude & Sig. ($\sigma$)\tnote{b} & $<$ DP\tnote{c} & $>$ DP \\
\hline
Global Mass\tnote{a} & 10.0 & 0.057$\pm$0.012 & 4.67 & 0.174$\pm$0.016 & 0.171$\pm$0.012\\
Local Mass & 9.4 & 0.037$\pm$0.012 & 3.17 & 0.161$\pm$0.012 & 0.183$\pm$0.014\\
\hline
Global U-R & 1.0 & 0.061$\pm$0.011 & 5.40 & 0.170$\pm$0.014 & 0.173$\pm$0.013\\
Local U-R & 1.1 & 0.053$\pm$0.011 & 4.70 & 0.170$\pm$0.013 & 0.173$\pm$0.013\\
\hline
\end{tabular}
\begin{tablenotes}
\item[a] Mass in $\log (\mstellar / M_\odot)$
\item[b] Significance is quadrature sum.
\item[c] DP refers to the \lq{Division Point}\rq\ location of the environmental property step. For example, \lq{$<$DP}\rq\ indicates the lower mass or bluer environments. 
\end{tablenotes}
\end{threeparttable}
\label{table:5yr_4values_BBC5D}
\end{center}
\end{table*}

As presented in \tref{table:c_split_steps_BBC5D}, when splitting the data based on $c$, we find that our three lost SNe all had $c > 0$. Interestingly, the step sizes remain fairly consistent for the $c < 0$ SNe Ia, but BBC5D exhibits noticeably smaller step sizes for $c > 0$ than BBC1D. This results in slightly less significant differences between red and blue SNe using BBC5D.  

\begin{table*}
\begin{center}
\caption{Subsample data when splitting the sample based on $c$ using a 5D bias correction.}
\begin{threeparttable}
\begin{tabular}{c c c c c c c} 
\hline
\multicolumn{2}{c}{Property} & \multicolumn{2}{c}{$c < 0$ Hubble Residual Step} & \multicolumn{2}{c}{$c > 0$ Hubble Residual Step} \\Name & Division Point & Magnitude & Sig. ($\sigma$)\tnote{b} & Magnitude & Sig. ($\sigma$) & Difference ($\sigma$)\tnote{c} \\
\hline
Number of Supernovae & & \multicolumn{2}{c}{306}& \multicolumn{2}{c}{366}&  \\
\hline
Global Mass\tnote{a} & 10.0 & $0.026\pm0.016$ & 1.7 & $0.088\pm0.018$ & 4.8 & 2.6\\
Local Mass & 9.4 & $0.010\pm0.015$ & 0.7 & $0.061\pm0.017$ & 3.6 & 2.2\\
\hline
Global U-R & 1.0 & $0.041\pm0.015$ & 2.8 & $0.081\pm0.017$ & 4.8 & 1.8\\
Local U-R & 1.1 & $0.026\pm0.015$ & 1.7 & $0.080\pm0.017$ & 4.8 & 2.4\\

\hline 
\end{tabular}
\begin{tablenotes}
\item[a] Mass in $\log (\mstellar / M_\odot)$
\item[b] Significance is quadrature sum.
\item[c] Difference is the quadrature sum difference in Hubble residual step magnitudes between red and blue subsamples.
\end{tablenotes}
\end{threeparttable}
\label{table:c_split_steps_BBC5D}
\end{center}
\end{table*}

\begin{figure*}
\begin{center}
\includegraphics[width=\linewidth]{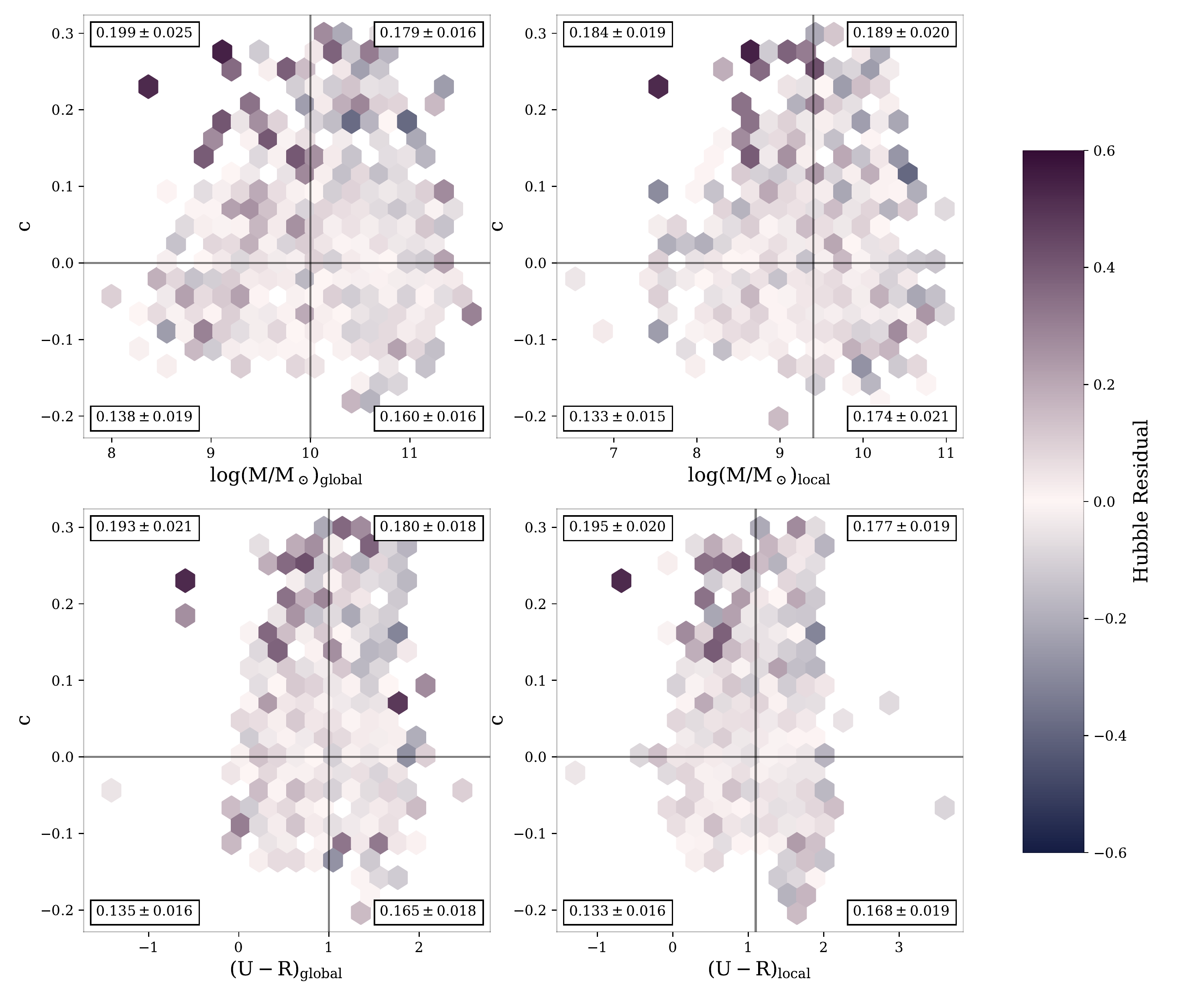}
\caption{As \fref{fig:quads}, but using a BBC5D bias correction.}
\label{fig:quads-BBC5D}
\end{center}
\end{figure*}

As above, the r.m.s. values across the board decreased using BBC5D (indicative of the increased scatter when only applying a redshift correction as in BBC1D), shown in \tref{table:c_split_rms_BBC5D}. As with the step sizes, the r.m.s. values for $c > 0$ decrease more, resulting in an smaller difference between the r.m.s. values between red and blue SNe using BBC5D. This is particularly noticeable when comparing the blue SNe in low $\mstellar$ or blue environments with those that are not, and is clear to see in \fref{fig:quads-BBC5D}. 

\begin{table*}
\begin{center}
\caption{Subsample r.m.s when splitting the sample based on $c$ using a 5D bias correction.}
\begin{threeparttable}
\begin{tabular}{c c c c c c} 
\hline
\multicolumn{2}{c}{Property} & \multicolumn{2}{c}{$c < 0$ Hubble Residual r.m.s.} & \multicolumn{2}{c}{$c > 0$ Hubble Residual r.m.s.} \\Name & Division Point & $< DP$\tnote{b} & $> DP$ & $< DP$ & $> DP$\\
\hline
Global Mass\tnote{a} & 10.0 & 0.138 $\pm$ 0.019 & 0.160 $\pm$ 0.016 & 0.199 $\pm$ 0.025 & 0.179 $\pm$ 0.016\\
Local Mass & 9.4 & 0.133 $\pm$ 0.015 & 0.174 $\pm$ 0.021 & 0.184 $\pm$ 0.019 & 0.189 $\pm$ 0.020\\
\hline
Global U-R & 1.0 & 0.135 $\pm$ 0.016 & 0.165 $\pm$ 0.018 & 0.193 $\pm$ 0.021 & 0.180 $\pm$ 0.018\\
Local U-R & 1.1 & 0.133 $\pm$ 0.016 & 0.168 $\pm$ 0.019 & 0.195 $\pm$ 0.020 & 0.177 $\pm$ 0.019\\

\hline 
\end{tabular}
\begin{tablenotes}
\item[a] Mass in $\log (\mstellar / M_\odot)$
\item[b] DP refers to the \lq{Division Point}\rq\ location of the environmental property step. For example, \lq{$<$DP}\rq\ indicates the lower mass or bluer environments. 
\end{tablenotes}
\end{threeparttable}
\label{table:c_split_rms_BBC5D}
\end{center}
\end{table*}

\begin{figure*}
\begin{center}
\begin{subfigure}[b]{0.49\textwidth}
    \centering
    \includegraphics[width=\linewidth]{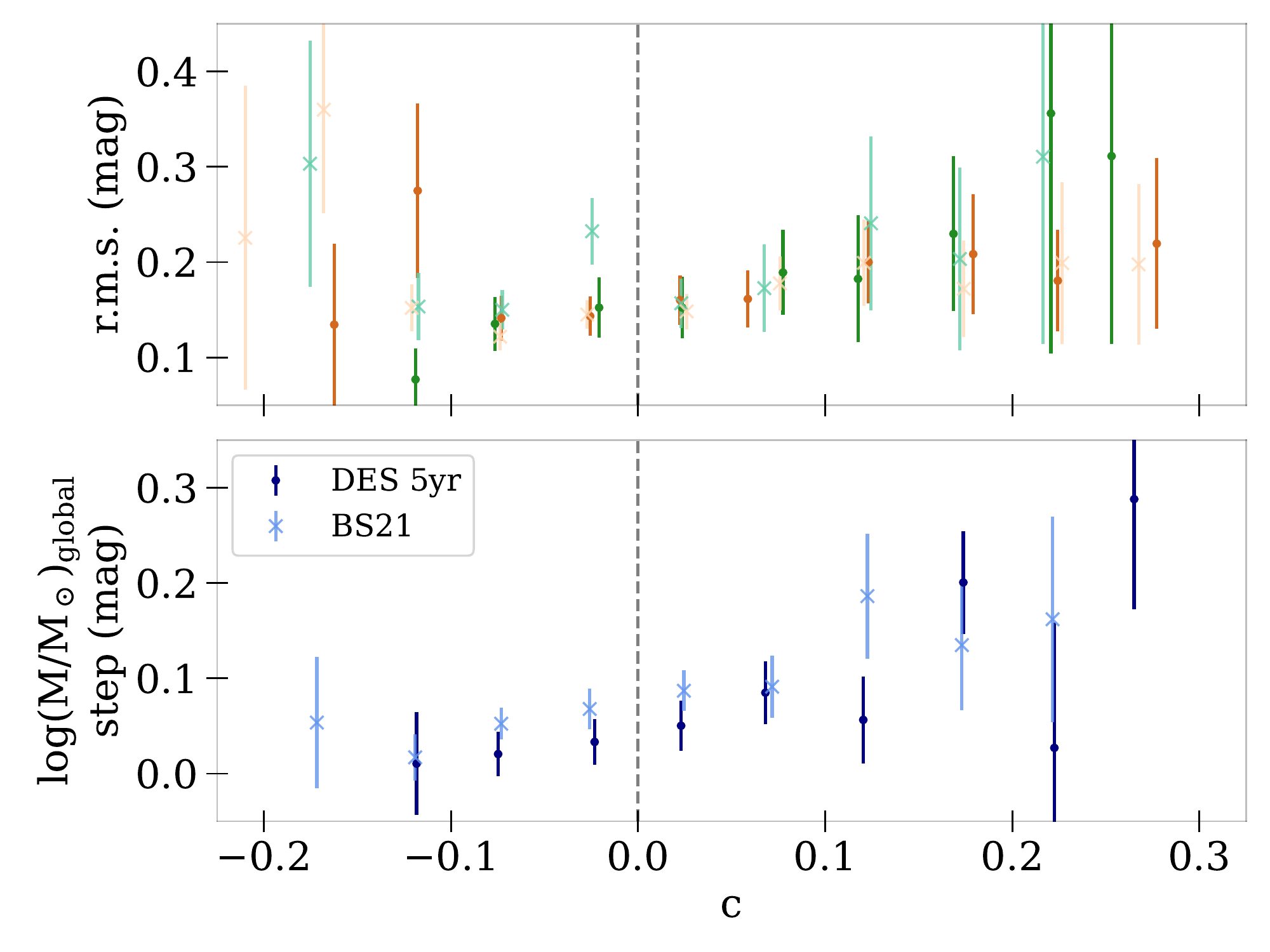}
    \caption{}
\end{subfigure}
\begin{subfigure}[b]{0.49\textwidth}
    \centering
    \includegraphics[width=\linewidth]{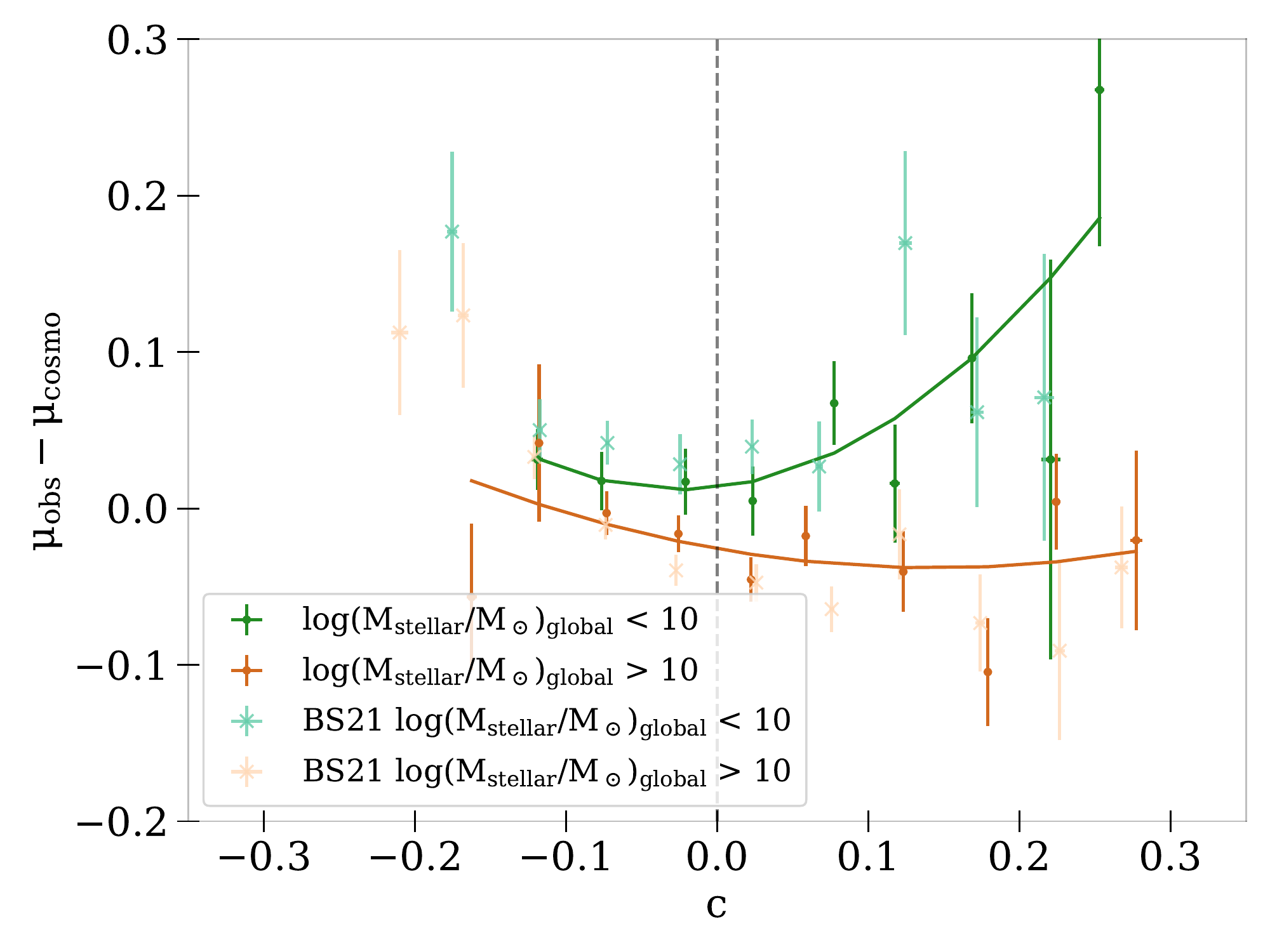}
    \caption{}
\end{subfigure}
\begin{subfigure}[b]{0.49\textwidth}
    \centering
    \includegraphics[width=\linewidth]{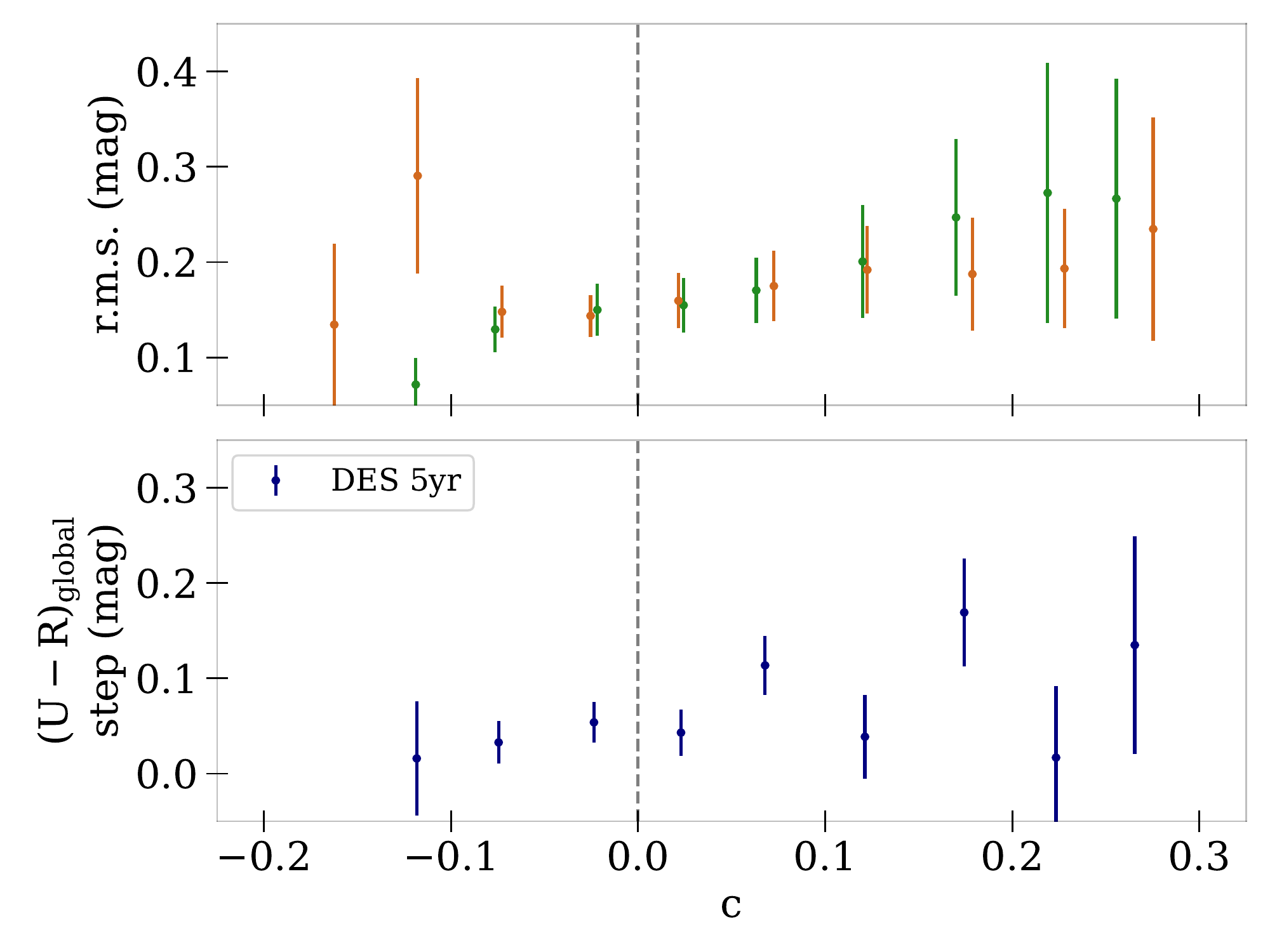}
    \caption{}
\end{subfigure}
\begin{subfigure}[b]{0.49\textwidth}
    \centering
    \includegraphics[width=\linewidth]{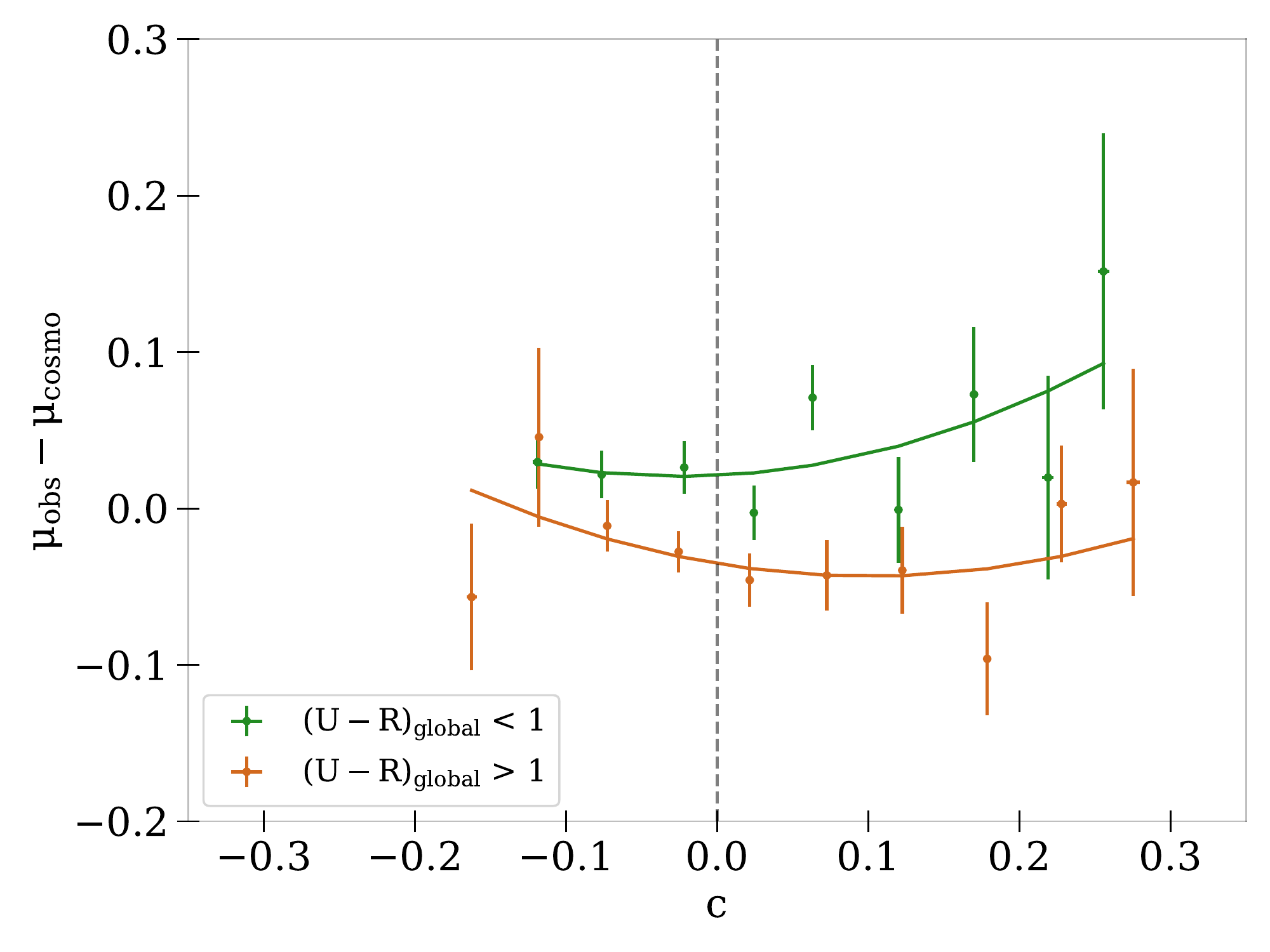}
    \caption{}
\end{subfigure}
\caption{As \fref{fig:likeBS20}, but using a BBC5D bias correction.}
\label{fig:likeBS20-BBC5D}
\end{center}
\end{figure*}

As in \sref{BS20}, we fit for trends between the observed environmental property $c$ dependent Hubble residual relationships. The resulting trends for global $\mstellar$ and $U-R$ are presented in \fref{fig:likeBS20-BBC5D}. As can be seen, the relationships are not identical to those using BBC1D, but they do follow the same general trends. 

Presented in \tref{table:fitremain_BBC5D} are the remaining environmental property Hubble residual steps when these $c$-dependent trends have been corrected for using the BBC5D bias correction. These remaining step values are fairly consistent with BBC1D, suggesting that our findings are not a result of the bias correction used. This agrees with \sref{results}, that a $c$-dependent global $U-R$ correction achieves the greatest reduction of remaining Hubble residual dispersion. Correcting for a $c$ - global $\mstellar$ still results in an intriguing $2\sigma$ remaining $U-R$ step. 

\begin{table*}
\begin{center}
\caption{Magnitudes and significances of remaining environmental property steps when fitting for relationships between $c$ and environmental properties, using a 5D bias correction.}
\begin{threeparttable}
\begin{tabular}{c c c c c c} 
\hline
\multicolumn{6}{c}{Fitting For:}\\
 & & Host Mass & Local Mass & Host U-R & Local U-R \\
\hline
\parbox[t]{2mm}{\multirow{8}{*}{\rotatebox[origin=c]{90}{Remaining Step In:}}} & {\multirow{2}{*}{Host Mass}} & 0.002$\pm$0.012 & $0.036\pm0.012$ & 0.013$\pm$0.012 & 0.021$\pm$0.012 \\
& & 0.1$\sigma$ & $3.0\sigma$ & 1.1$\sigma$ & 1.7$\sigma$\\
& {\multirow{2}{*}{Local Mass}} & $0.007\pm0.011$ & $0.000\pm0.011$ & $0.005\pm0.011$ & $0.004\pm0.011$ \\
& & $0.6\sigma$ & $0.0\sigma$ & $0.4\sigma$ & $0.4\sigma$\\
& {\multirow{2}{*}{Host U-R}} & 0.023$\pm$0.011 & $0.041\pm0.011$ & 0.001$\pm$0.011 & 0.017$\pm$0.011\\
& & 2.0$\sigma$ & $3.7\sigma$ & 0.1$\sigma$ & 1.5$\sigma$\\
& {\multirow{2}{*}{Local U-R}} & 0.018$\pm$0.011 & $0.032\pm0.011$ & 0.005$\pm$0.011 & 0.001$\pm$0.011\\
& & 1.6$\sigma$ & $2.8\sigma$ & 0.4$\sigma$ & 0.1$\sigma$\\
\hline 
\end{tabular}
\begin{tablenotes}
\item All steps given in mag.
\item Step division points as described in Section~\ref{csplit}.
\end{tablenotes}
\end{threeparttable}
\label{table:fitremain_BBC5D}
\end{center}
\end{table*}

\section{Different Classifiers and Training Sets} \label{diff_class}
For this analysis, we have focused on using the \textsc{SuperNNova} (SNN) \citep{Moller2020} photometric classifier trained on core-collapse templates from \citet{Vincenzi2019}, requiring $P(Ia) > 0.5$ as this is currently the preferred choice for the final DES5YR sample within the DES-SN collaboration. However, we repeated this analysis using various combinations of classifier, templates and probability cuts. These are defined as follows:
\begin{enumerate}
    \item SNN trained on \citet{Vincenzi2019} templates, $P(Ia) > 0.5$
    \item SNN trained on \citet{Vincenzi2019} templates, $P(Ia) > 0.8$
    \item SNN trained on \citet{Vincenzi2019} templates, $P(Ia) > 0.95$
    \item SNN trained on \citet{Jones2017} templates, $P(Ia) > 0.5$
    \item SNN trained on \citet{HS21} templates, $P(Ia) > 0.5$
    \item Supernova Identification with Random Forest \citep[SNIRF; an extension of][]{Dai2018} trained on \citet{Vincenzi2019} templates, $P(Ia) > 0.5$
    \item SNIRF trained on \citet{Jones2017} templates, $P(Ia) > 0.5$
\end{enumerate}

The overwhelming majority of candidate objects have a high probability of being a SNe Ia of $P(Ia) > 0.95$. When the different $P(Ia)$ samples undergo the additional quality cuts that are specific to this analysis, the final sample sizes were comparable (only changing by a few objects), and there was no clear difference in results for each sample. In other words, the objects that had a low $P(Ia)$ typically also had the largest uncertainties in $x_1$, $c$, and environmental properties, meaning that they were removed in each case. As investigated in \citet{Vincenzi2020} for DES, core-collapse contamination using each of the above training sets with SNN is low, with a maximum of 3.5 per cent of the sample consisting of potential core-collapse SNe.


\bsp	
\label{lastpage}
\end{document}